\documentclass[aps,nofootinbib,twocolumn,superscriptaddress,letterpaper,amsmath,amssymb]{revtex4}

\usepackage{hyperref}
\usepackage{graphicx}
\usepackage{dcolumn}
\usepackage{bm}
\usepackage{amssymb}
\usepackage{feynmf}
\usepackage{slashed}
\usepackage{multirow}
\usepackage{color}
\usepackage{array}
\usepackage{url}
\usepackage{tikz}
\usepackage{siunitx}
\usepackage[percent]{overpic}

\begin{document}

\title{Learning to Isolate Muons in Data}

\author{Edmund Witkowski}
\email{witkowse@uci.edu}
\affiliation{Department of Physics and Astronomy, University of California, Irvine, CA 92697}
  
  \author{Benjamin Nachman}
\email{bpnachman@lbl.gov}
\affiliation{Physics Division, Lawrence Berkeley National Laboratory, Berkeley, CA 94720, USA}
\affiliation{Berkeley Institute for Data Science, University of California, Berkeley, CA 94720, USA}

\author{Daniel Whiteson}
\email{daniel@uci.edu}
\affiliation{Department of Physics and Astronomy, University of
  California, Irvine, CA 92697}

\begin{abstract}
We use unlabeled collision data and weakly-supervised learning to train models which can distinguish prompt muons from non-prompt muons using patterns of low-level particle activity in the vicinity of the muon, and interpret the models in the space of energy flow polynomials. Particle activity associated with muons is a valuable tool for identifying prompt muons, those due to heavy boson decay, from muons produced in the decay of heavy flavor jets. The high-dimensional information is typically reduced to a single scalar quantity, isolation, but previous work in simulated samples suggests that valuable discriminating information is lost in this reduction. We extend these studies in LHC collisions recorded by the CMS experiment, where true class labels are not available, requiring the use of the invariant mass spectrum to obtain macroscopic sample information. This allows us to employ Classification Without Labels (CWoLa), a weakly supervised learning technique, to train models. Our results confirm that isolation does not describe events as well as the full low-level calorimeter information, and we are able to identify single energy flow polynomials capable of closing the performance gap.  These polynomials are not the same ones derived from simulation, highlighting the importance of training directly on data. 
\end{abstract}

\date{\today}

\maketitle

\section{Introduction}

Data collected in hadronic collisions offer a significant opportunity to precisely test the Standard Model (SM) and to search for physics beyond the SM (BSM). The identification of muons resulting from electroweak boson decays (called `prompt') is a crucial part of many such studies, as muons are typically well measured and have low rates of background. An important source of background for these events comes from muons produced within jets from decays in flight. This `non-prompt' background is largest at the lower end of the muon transverse momentum spectrum, which has become important in searches for supersymmetry~\cite{Aaboud:2017leg,SCHOFBECK2016631,Khachatryan:2015kxa, ATLAS:2022hbt,CMS:2021edw,ATLAS:2019lng} as well as for low-mass resonances~\cite{Hoenig:2014,ATLAS:2023vxg,LHCb:2020ysn,CMS:2021pcy}. 

 Prompt muons tend to have less nearby detector activity as compared to muons from jets, which are found near hadrons from the rest of the jet. The concept of {\it isolation} is therefore important to much of the work involving the discrimination of prompt muons from the non-prompt backgrounds. A complete description of the isolation requires capturing the high-dimensional data in the vicinity of the muon. In practice, high-dimensional data are challenging to analyze and isolation is typically reduced to a scalar quantity~\cite{Sirunyan:2017ulk,CMS:2017yfk,ATLAS:2020auj}
 However, in the reduction from a high-dimensional (low-level) representation of the data to a lower-dimensional (high-level) one, information can be lost. 
 
Deep learning with low-level inputs has been demonstrated to exceed the performance of engineered high-level observables on a number of tasks in high energy physics, starting with Refs.~\cite{Baldi:2014kfa,deOliveira:2015xxd} and now including many studies~\cite{Feickert:2021ajf}.  In the context of prompt muon identification, deep neural networks were able to outperform classical isolation definitions using simulated data -- by as much as 50\% in non-prompt background rejection at a prompt muon efficiency of 50\%~\cite{Collado:2020ehf}.  This was achieved by processing all of the calorimeter cells\footnote{The previous work mentioned here only used calorimeter information, though this study considers both calorimeter and track information.} in the vicinity of the muon, corresponding to roughly 1000 dimensions per event.  Significant suppression of non-prompt backgrounds with a deep learning approach has the potential to improve the precision and sensitivity of many measurements and searches involving muons at the Large Hadron Collider (LHC).

However, previous studies were based on simulations, with relatively simple detector effects.  Hadronic final states are complex and difficult to model, so it is reasonable to be concerned that the performance of a deep learning-based isolation strategy trained on simulated events may depend on details of the simulation which are not faithful reproductions of collider data.  Scale factors derived using standard tag-and-probe methods~\cite{ATLAS:2016lqx,CMS:2018rym}  may correct the efficiency, but the performance in data would be suboptimal~\cite{Cesarotti:2019nax}.  Achieving optimal performance in data  requires training with data.  The limitation is that data are not labeled as prompt or not-prompt, so the {\it supervised} machine learning strategies used in previous studies and which require such labels cannot be applied to data.

We propose to overcome this limitation with {\it weakly supervised} learning.  In contrast to supervised learning, where every event is labeled with certainty as prompt or non prompt, weakly supervised learning is trained with noisy labels, which describe the overall composition of the sample but not individual events.  Specifically, we use the Classification Without Labels (CWoLa)~\cite{Metodiev:2017vrx} approach to weak supervision where two samples of training events are prepared.  One sample is dominated by prompt muons, and receives the noisy label of `signal' (and will be called `prompt abundant'); the second sample, while still mostly containing prompt muons, has a relatively higher fraction of non-prompt muons and receives the noisy label of `background' (and will be called `prompt moderate'). Under mild assumptions, training a standard classifier with these noisy labels converges to the same classifier found in a supervised setting.  While weak supervision has been used previously for data analysis~\cite{CMS:2019eih,Mikuni:2781479,Collins_2019,Collins:2018epr,ATLAS:2020iwa}, these studies only used 2-18 inputs.  Our goal is to approach the muon isolation problem with weak supervision directly on low-level, high-dimensional ($\mathcal{O}(100)$) inputs. While the inputs are high dimensional enough to hold a large number of detected objects, this is only necessary for a small number of events, as on average the inputs have $\mathcal{O}(10)$ non-zero entries.

Even if proven effective in data, deep networks operating on low-level observables can be opaque. To improve the interpretability and compactness of the network, we follow Ref.~\cite{Collado:2020ehf}, bridging the performance gap between the low-level observables and classical isolation variables through a small set of additional high-level observables identified by the decisions of a network operating at the low-level. We search for new high-level observables among the Energy Flow Polynomials (EFP)~\cite{Komiske:2017aww}, a set of relatively simple combinations of energies and angles of reconstructed objects within the isolation cone.  EFP observables are  identified automatically using the Average Decision Ordering (ADO) method~\cite{Faucett:2020vbu}, which uses the decisions of the low-level network as a guide.  While still complex, the resulting EFP is more physically interpretable than the original deep neural network.  Interestingly, the first EFP selected through this process was not identified in the previous study as a top candidate for closing the corresponding gap in simulation~\cite{Collado:2020ehf}.  This is one more reason why it is essential here to train directly on data.

This paper is organized as follows.  Section~\ref{sec:data} introduces the dataset, which is from the CMS experiment~\cite{CMS:2008xjf,cms-open-data}.  Then, Sec.~\ref{sec:methods} describes the machine learning strategy.  Numerical results are presented in Sec.~\ref{sec:results}.  The paper ends with conclusions and outlook in Sec.~\ref{sec:conclusions}.

\section{Dataset}
\label{sec:data}

Proton-proton collisions at $\sqrt{s}=$ 8~TeV were 
 recorded in 2012 and curated by the CMS Collaboration and made available through the CERN Open Data Portal~\cite{cms-open-data}.  The number of collisions corresponds to 19.5~fb$^{-1}$.  Reconstruction was performed with the Particle Flow (PF) algorithm~\cite{CMS:2017yfk}, which integrates calorimeter and tracker information to approximate individual particle four-vectors.  The PF algorithm also assigns a particle identification (PID) from one of the following types: muon, charged hadron, neutral hadron, photon, or pileup.  For the charged PF objects, the sign of the charge is reconstructed.  PF object momenta are represented by their transverse momentum ($p_{\mathrm{T}}$), pseudorapidity ($\eta$), and azimuthal angle ($\phi$).

\begin{figure}
    \centering
    \includegraphics[width=0.90\linewidth]{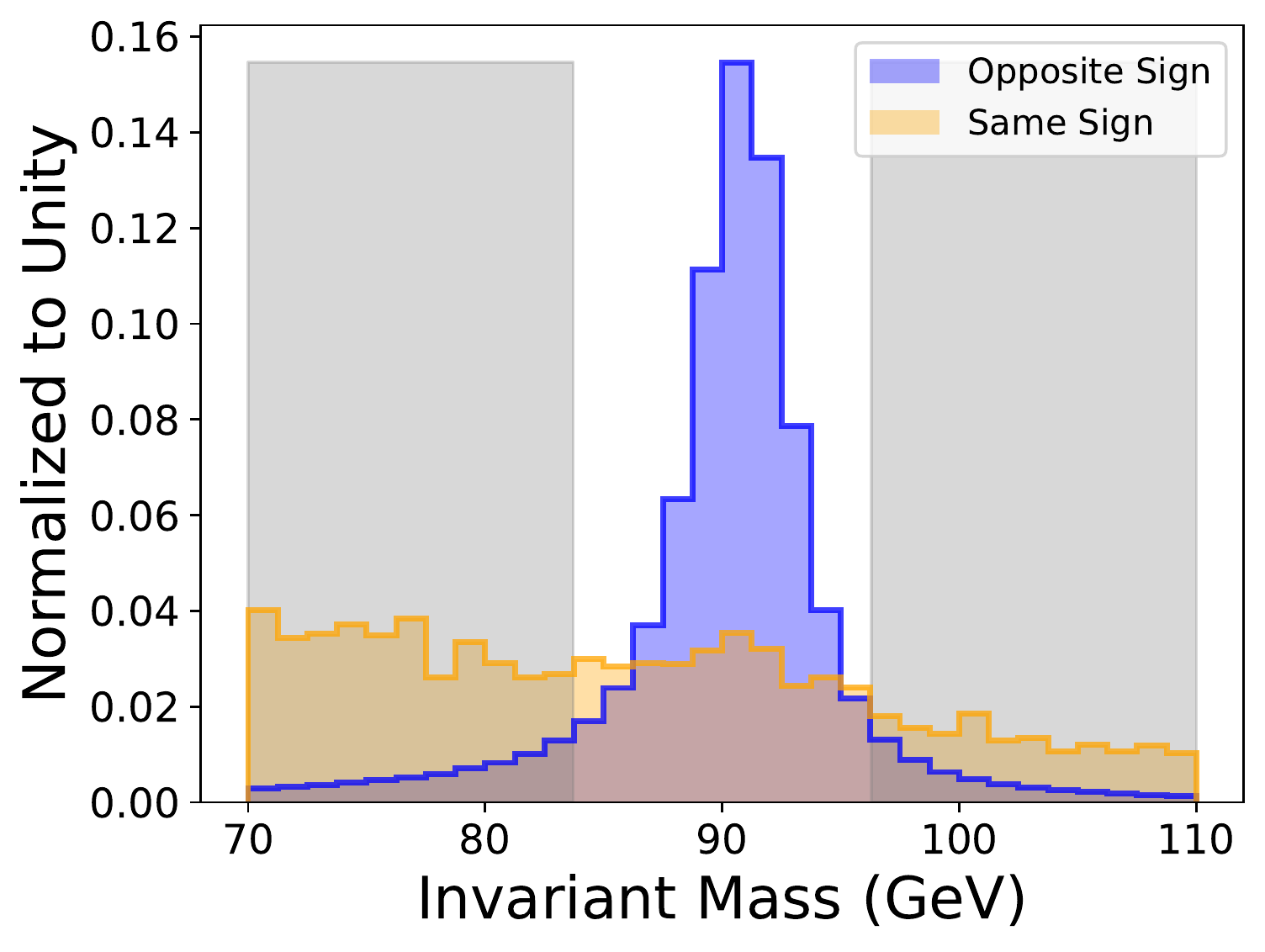}
    \caption{Histogram of the dimuon invariant mass near the $Z$ boson peak, for events in data with identical (yellow) or opposite (blue) electric charges. The unshaded area indicates the region for the oppositely-charge pairs which comprises our ``prompt muon abundant" sample. The grey, shaded region for the oppositely-charge pairs, as well as the entire region for identically-charged pairs, comprise our ``prompt muon moderate" sample.}
    \label{fig:z_peak}
\end{figure}

We select events with exactly two muons,  both with $p_{\mathrm{T}} \geq 25$ GeV, $|\eta| \leq 2.1$, and with a dimuon invariant mass between $70$ and $110$ GeV to accommodate the $Z$ boson mass of 90 GeV~\cite{ParticleDataGroup:2022pth}. Events are separated into two samples which have different mixtures of prompt and non-prompt muon events, as is required by the CWoLa method. 
One sample, with a higher fraction of non-prompt muons, consists of all events in which the muons have identical electric charge, as well as events with muon pairs of opposite electric charge but reconstructed invariant mass far from the $Z$ boson invariant mass, below $84$ GeV or above $96$ GeV. This sample is referred to as the ``prompt muon moderate sample."  The remaining events, which are almost entirely prompt muons, form the complementary sample and are referred to as the ``prompt muon abundant sample."  These regions are illustrated in Fig~\ref{fig:z_peak}.  The opposite sign sample is almost entirely from $Z$ boson decays and so is peaked at the $Z$ boson mass.  The same sign sample is mostly from decays in flight and has a nearly smooth and steeply falling spectrum. 

\begin{figure}
\includegraphics[width=0.90\linewidth]{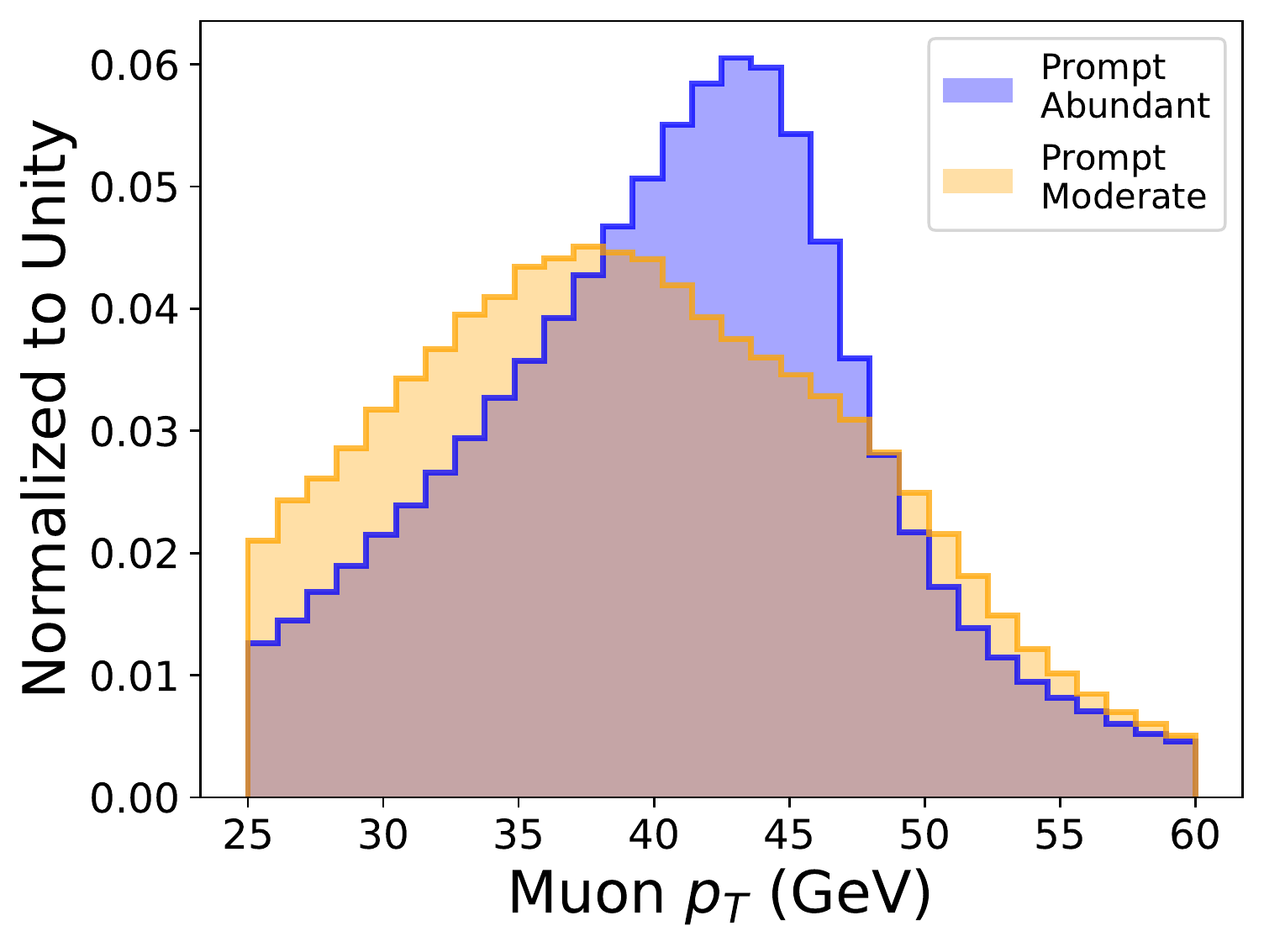}
\includegraphics[width=0.90\linewidth]{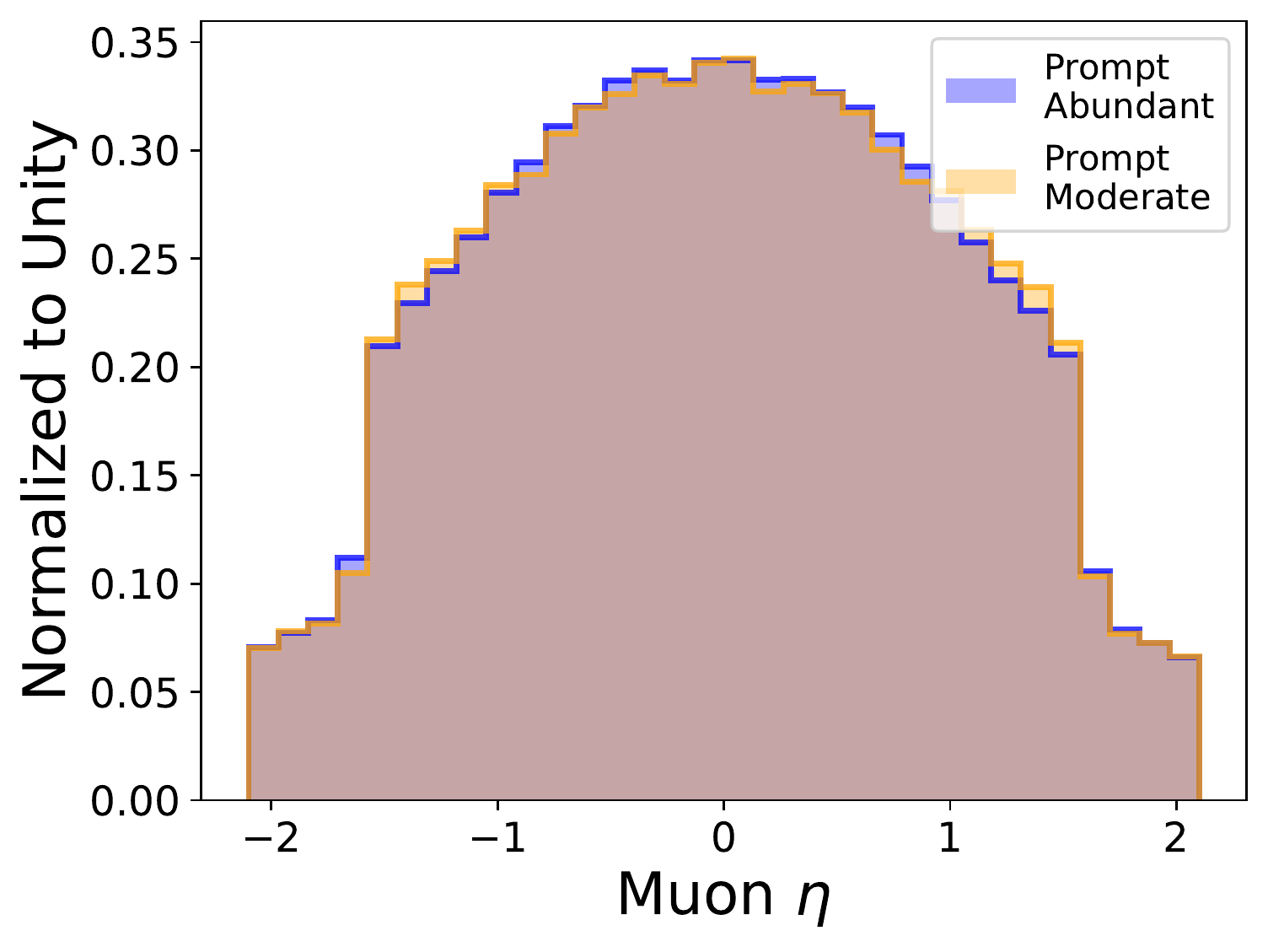}
\caption {Histograms of muon $p_{\mathrm{T}}$ and pseudorapidity $\eta$ in the two samples with varying fractions of prompt muons, as defined in text and Fig.~\ref{fig:z_peak}.}
\label{fig:muon_dists}
\end{figure}

In order to ensure that the two samples have similar kinematic distributions, event weights are computed so that the muon $p_{\mathrm{T}}$ and $\eta$ spectra are the same between the prompt-enriched and non-prompt-enriched samples.   The unbinned likelihood ratio is estimated using a two-dimensional Kernel Density Estimator with Gaussian kernels.  The pre-weighted spectra are displayed in Fig.~\ref{fig:muon_dists}.  The $p_{\mathrm{T}}$ spectrum is peaked near $m_Z/2$ and the sharp features in the muon histogram are due to detector acceptance effects.  We additionally validate the core assumption of CWoLa (see Sec.~\ref{sec:methods}) -- that the (non)prompt muons look the same in both samples -- using samples of simulated muons; see Appendix~\ref{sec:app}.

\begin{figure}
\includegraphics[width=0.95\linewidth]{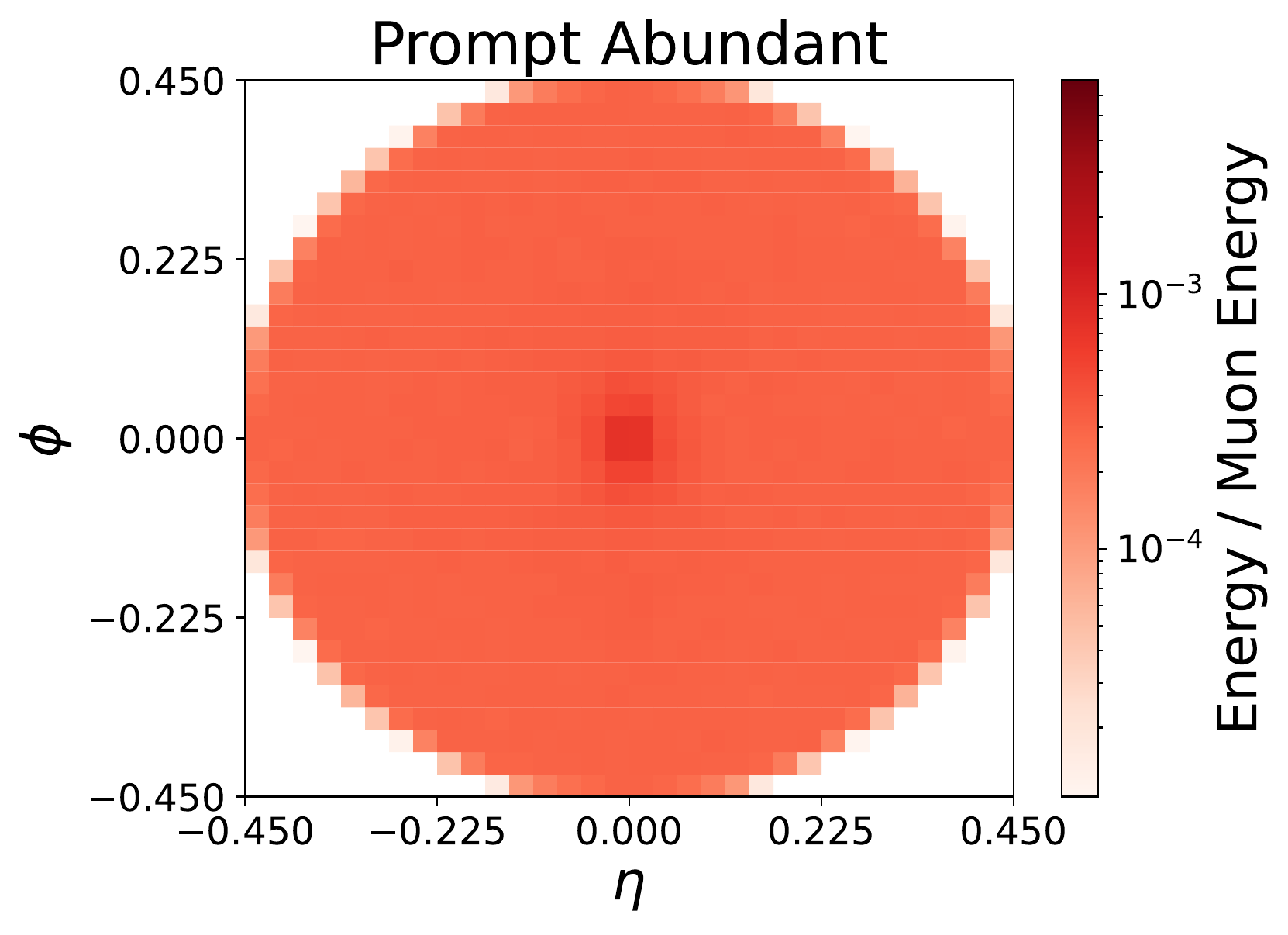}
\includegraphics[width=0.95\linewidth]{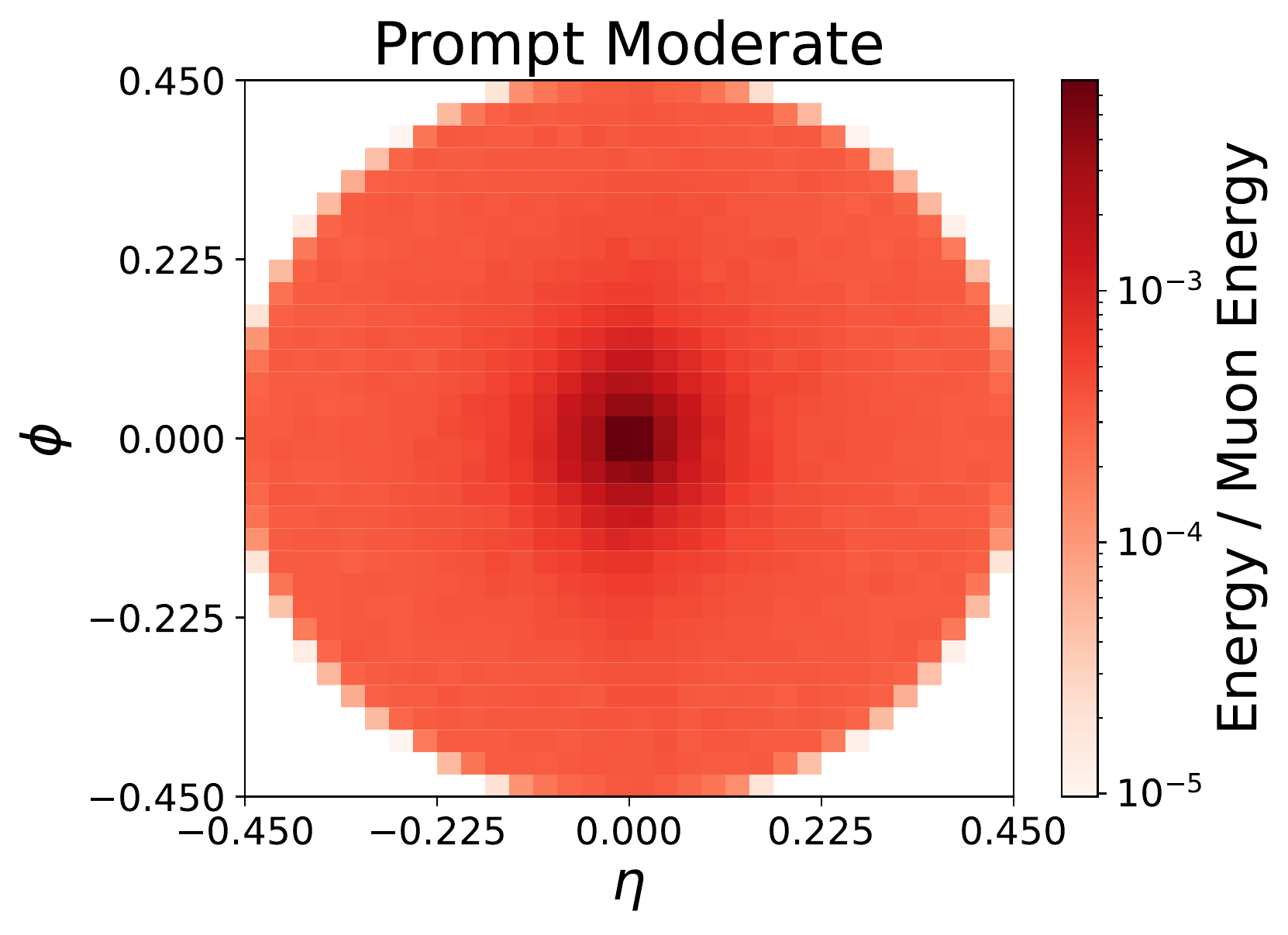}
\caption {The average image of hadronic activity in the vicinity of an identified muon, in angular coordinates of azimuthal angle $\phi$ and pseudorapidity $\eta$, for our two training samples, one which is dominated by prompt muons (top) and a second which has a more moderate mixture of prompt and non-prompt muons (bottom). The muon itself is excluded from these visualizations, but the energies are normalized by that of the muon.}
\label{fig:average_images}
\end{figure}

Once events are selected, they are formatted to be used as inputs to the neural networks. The low-level inputs are comprised of the $p_{\mathrm{T}}$, $\eta$, $\phi$, and PID for each constituent within a 0.45 radius around a given muon.  We additionally preprocess the low-level input by centering on the muon and dividing the momenta by the muon transverse momentum.  A visualization of the momentum in the vicinity of the muon, not including the muon itself, for both samples is shown in Fig.~\ref{fig:average_images}. We see that the sample means per pixel have distinct distributions, with the more prompt sample being  more uniform. 

\begin{figure}
    \centering
    \includegraphics[width=0.90\linewidth]{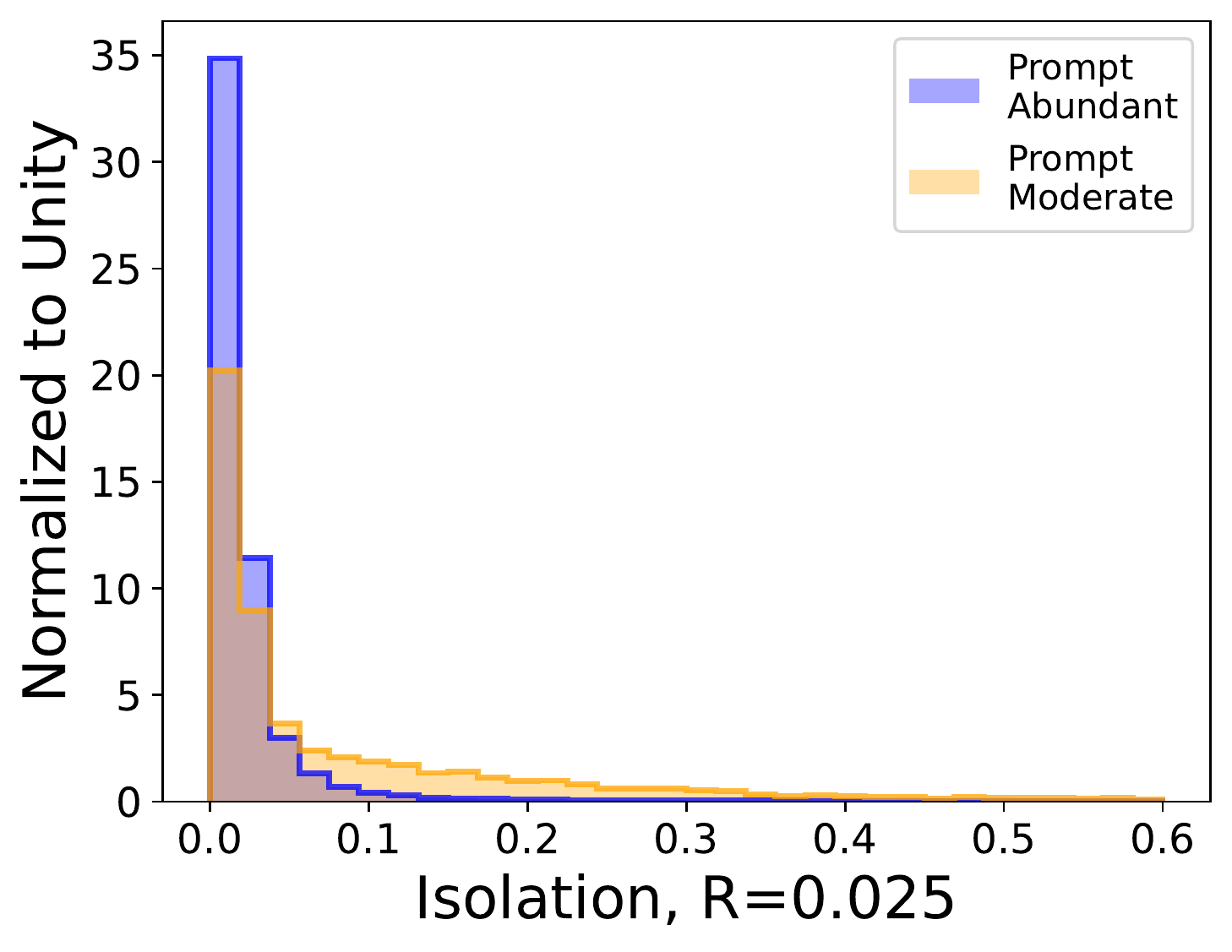}
    \includegraphics[width=0.90\linewidth]{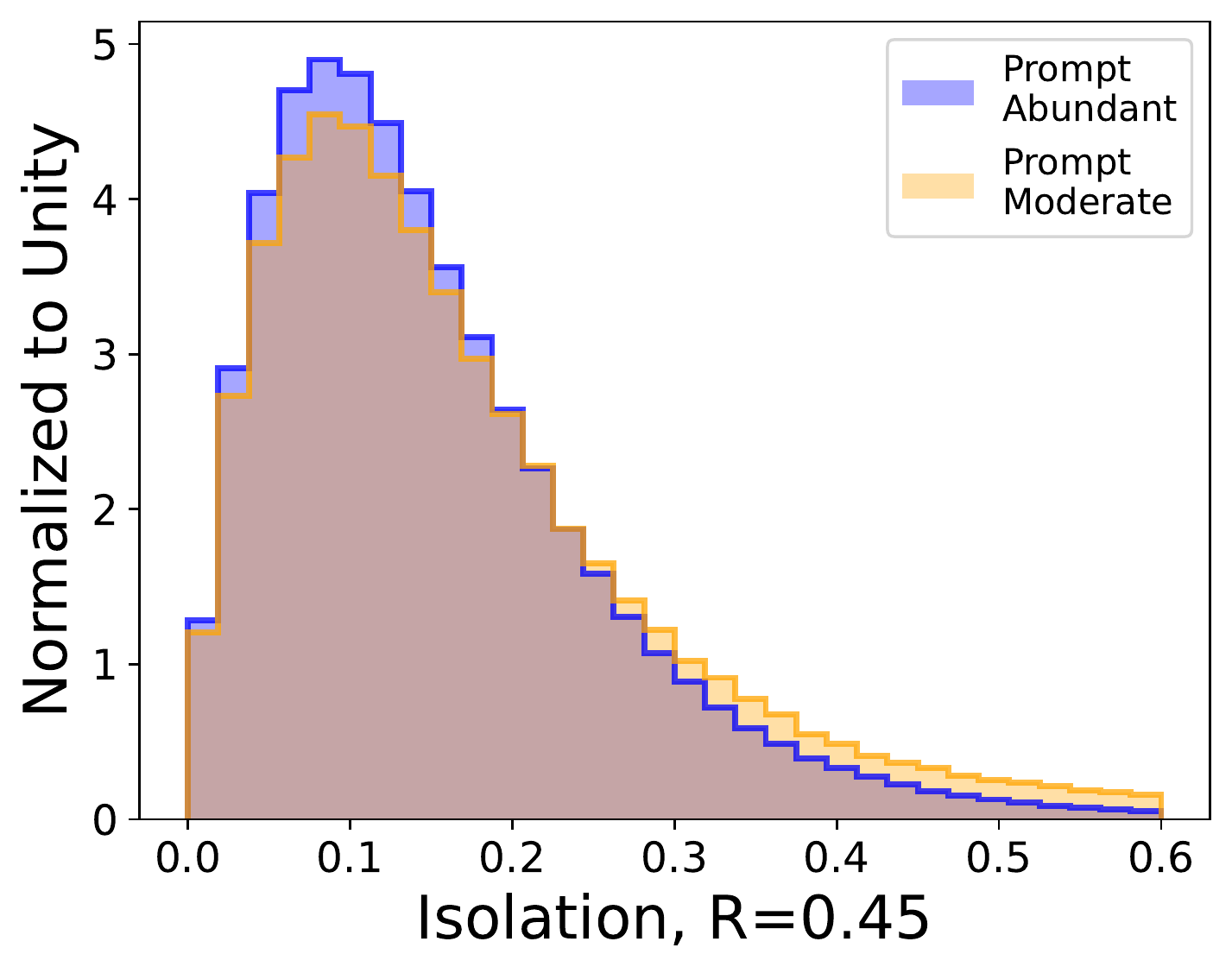}
    \caption{Histograms of the muon isolation (defined in Eq.~\ref{eq:isolation_eq}) for each of our training samples, one of which is dominated by prompt muons, for two choices of isolation cone radius parameter $R_0=0.025$ (top) and $R_0= 0.45$ (bottom).}
    \label{fig:iso}
\end{figure}

Traditional, high-level scalar observables are calculated from the low-level data. These observables include the summed $p_{\mathrm{T}}$ of non-muon objects in an event, isolation, and EFP observables. We calculate isolation as defined in Eq.~\ref{eq:isolation_eq}, where $\textrm{h}_\pm$ and $\textrm{h}_0$ denote charged and neutral hadrons, respectively. This definition quantifies the activity around a muon within a given radius strictly in terms of Particle Flow objects and treats the objects differently according to their Particle Flow ID. The expression is composed of terms which sum over the transverse momenta of the non-muon Particle Flow candidates within the chosen radius, and the result is normalized by the muon momentum. Pileup is mitigated by subtracting half of its sum from the neutral hadron and photon sums, and clamping the result of this subtraction at 0. Distributions of the isolation for two choices of cone radius are shown in Fig~\ref{fig:iso}. The larger of the two choices of radius tends to yield larger isolation values, as one might expect.

\begin{multline}
   \label{eq:isolation_eq}
   I_\mu(R_0) = \biggl[\sum_{i, R<R_0}^{N_{\textrm{h}_\pm}} p_{\textrm{T},\textrm{h}_\pm}^{\textrm{i}} + \textrm{max}\biggl(0, \sum_{i, R<R_0}^{N_{\textrm{h}_0}} p_{\textrm{T},\textrm{h}_0}^{\textrm{i}} \\ 
   + \sum_{i, R<R_0}^{N_\gamma} p_{\textrm{T},\gamma}^{\textrm{i}} - \frac{1}{2}\sum_{i, R<R_0}^{N_{\textrm{pileup}}}p_{\textrm{T},\textrm{pileup}}^{\textrm{i}}\biggr)\biggr] / p_{\textrm{T},\textrm{muon}}
\end{multline}

We calculate isolation quantities for a set of radii from $0.025$ - $0.45$ in steps of $0.025$. CMS has previously studied isolation at radius of $0.3$~\cite{CMS:2017yfk}, which is included in our generated set.

While in principle the demonstration of weak supervision as a technique for learning to improve muon isolation beyond cone-based quantities could use simulation instead of data, we have chosen to use collider data for a number of reasons.  First, realistic simulation of muon isolation is very challenging, for both the prompt and non-prompt categories; see App.~\ref{sec:app}. Second, a demonstration in data can confirm (or refute) the results of earlier studies in simulation, which showed a significant gap between the power of isolation cones and full use of the lower-level data.  If such a gap exists in collider data, it would indicate that additional information is available in nature; the interpretation of that gap in terms of EFP observables will provide clues as to the physical processes involved, and the size of the gap can motivate a further study in a complete experimental context.  For this reason, we also do not estimate systematic uncertainties, which would be required before application to searches and measurements.  As a data-driven method, there are no simulation-based uncertainties, but there would be method closure uncertainties related to the underlying assumptions of CWoLa and sPlots.

\section{Methods}
\label{sec:methods}

Classification Without Labels (CWoLa) defines a weakly supervised setting which relies on the principle that given two classes, an optimal classifier may be obtained by training to discriminate between two samples composed of different mixtures of the classes, rather than training directly on two pure class samples. This technique only requires that the two samples have different class mixtures, and these mixtures do not need to be known in order for training to proceed.  The essential assumption is that class fraction is the only feature that determines the different properties of the two samples.  This means that the spectrum of radiation around the muon for prompt leptons is identical for the prompt muon abundant and the prompt muon moderate samples. Similarly, the probability density for hadrons around the muon for non-prompt leptons should be the same within the prompt muon abundant and the prompt muon moderate samples.  We expect this to be the case here, since the invariant mass of the muons and their relative electric charges should statistically independent from the radiation pattern around the muons given the prompt status.  This expectation is validated in simulation in App.~\ref{sec:app}.

\begin{figure}
\includegraphics[width=0.90\linewidth]{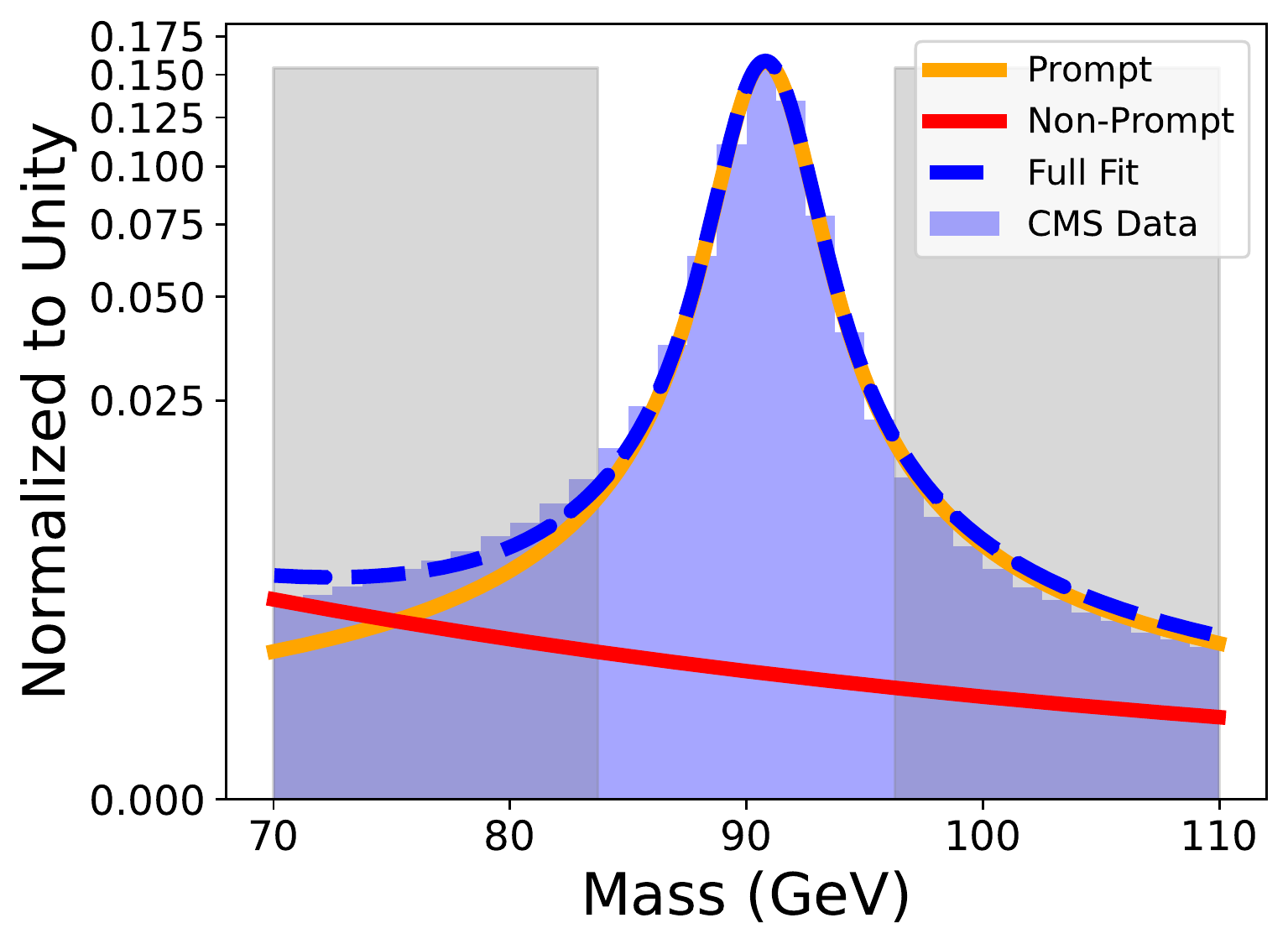}
\caption {A visualization of the masses overlaid with the fit and its prompt / non-prompt components. The shaded regions indicate events which are included in the relatively less prompt sample. Here we fit the full CMS sample used in the study, finding that it is $95.6 \pm 0.6\%$ prompt overall.}
\label{fig:mass_fit}
\end{figure}

While CWoLa does not need class labels to derive a classifier, some class information is required to determine the performance of the method.  The only information needed is the proportion of prompt muons in each sample; from this information, it is possible to characterize the full tradeoff between signal efficiency and background rejection.   The prompt-muon fraction is measured directly from the data in each sample by modeling the invariant mass distribution as a mixture model with two components: one peaking component of $Z$ bosons which decay to two prompt muons, and a second, non-peaking component. The invariant mass spectrum  is fit using a Voigt profile and an exponential function for the respective components. Fitting is done with Scipy v1.7.3~\cite{2020SciPy-NMeth} and visually demonstrated in Fig~\ref{fig:mass_fit}, where the fit is applied to the full dataset, finding an overall prompt fraction of $95.6 \pm 0.6\%$, where the error bar corresponds to $1\sigma$ statistical. In the ``prompt muon abundant" sample, the prompt fraction is measured to be $98.9$\%; in the ``prompt muon moderate'' sample, the prompt fraction is measured to be $56.0$\%. This is the first application of weak supervision in particle physics where the relative proportions have also been extracted directly from data.

Characterizing the network performance is non-trivial without pure samples. To measure the efficiency of a varying network threshold in the prompt and non-prompt samples, one could fit the distribution of the invariant mass of events surpassing each threshold.  Measurement of the efficiencies of each class allows calculation of performance metrics, such as the standard Receiver Operating Characteristic (ROC) and its associated statistics. However, fits are expensive and stochastic. Fitting the mass spectrum for each threshold output can be avoided using the sPlots technique~\cite{Pivk:2004ty}, which can decompose the prompt and non-prompt contributions to distributions of the network output given weights from the single invariant mass fit into the full sample. sPlots assumes that the variable being weighted is statistically independent of the invariant mass, within the individual classes. This is approximately true for our discriminating variables, such as model outputs, and so the method can be applied. Once the variable has been separated by the components, the resulting histograms may be integrated to calculate true and false positive rates, and construct a ROC curve. Performance is evaluated through the Area Under the Curve (AUC) and the signal efficiency at 50\% background efficiency.  While we do not perform a full determination of the uncertainty, we do consider statistical sources of uncertainty from the training and from the fit\footnote{While these are the only sources of uncertainty quantified in Table~\ref{table:perf_summary}, other sources are present, such as a bias due to imperfect description of the mass distribution by the fit function.}. While not an uncertainty per se~\cite{Nachman:2019dol}, the statistical variation from the finite size of the training dataset\footnote{The random initialization of the network is also folded into this estimation.} gives a sense for the stability and optimality of the result.  This effect is estimated using bootstrapping~\cite{10.1214/aos/1176344552} with 100 event ensembles with a new classifier trained per ensemble.  Additionally, we propagate the statistical uncertainty from the fit in each ensemble by sampling 100 times from the fitted parameter covariance matrix. Metrics are recomputed and averaged across each ensemble, and we report the $1\sigma$ confidence intervals according to the resulting set of values. 

We consider two types of neural networks: high-level networks with an increasing list of engineered observables (such as isolation) and low-level networks that process the full muon image.  For the high-level networks, one of our goals is to determine the minimal set of isolation observables that will saturate the performance.  To do this, we start by training a network using the single isolation cone corresponding to the largest radius in our set and subsequently train networks with incrementally smaller cones included as inputs. The summed event $p_{\mathrm{T}}$ is included as an input in all of these sets, in order to be sensitive to overall normalization effects. 

The low-level networks take the full, high-dimensional representations of the events as inputs. We use the deep sets architecture~\cite{zaheer2018deep} implemented as Particle Flow Networks (PFNs)~\cite{Komiske:2018cqr} to process these data.  This architecture was chosen because the inputs are a permutation-invariant, variable-length set of four-vectors and so a point-cloud model is the natural choice for processing them.  Deep sets models are composed of two fully connected networks. The first network embeds each particle flow object (represented by $(p_{\mathrm{T}},\eta,\phi,\mbox{PID})$) into a latent space.  The second network processes the sum of these latent space vectors across all inputs. The sum operation is permutation invariant and can readily process variable-length inputs.

Additionally, we strive to close the gap in performance between low- and high-level networks using relatively simple variables.  Energy Flow Polynomials (EFPs)~\cite{Komiske:2017aww} serve as a set of potential variables for this purpose.  EFPs are a set of parameterized functions which sum over objects within an event, were each term is weighted using the angular relations between these objects. EFPs can be represented using graphs, where

\begin{align}
	\text{each node} &\Rightarrow \sum_{i = 1}^N z_i, \label{eq:EFP_node}  \\
	\text{each $k$-fold edge} &\Rightarrow \left(\theta_{ij}\right)^k \label{eq:EFP_edge} . 
\end{align}
\begin{align}
	(z_i)^\kappa &= \left(\frac{p_{\textrm{T}i}}{\sum_j p_{\textrm{T}j}} \right)^\kappa, \label{eq:EFP_z} \\
	\theta^\beta_{ij} &= \left(\Delta \eta_{ij}^2 + \Delta \phi_{ij}^2 \right)^{\beta/2}. \label{eq:EFP_Theta}
\end{align}

When $\kappa=1$ the EFPs form a basis for Infrared and Collinear (IRC)-safe observables~\cite{Komiske:2017aww}. We compute a set of EFPs which contains IRC-safe, as well as unsafe, information, using the same parameterizations as in Ref.~\cite{Collado:2020ehf}: $\kappa \in [-1,0,\frac{1}{4},\frac{1}{2},1,2]$ and $\beta \in [\frac{1}{4},\frac{1}{2},1,2,3,4]$, for graphs with up to $n=7$ nodes and up to $d=7$ edges. 

We use the Average Decision Ordering (ADO)~\cite{Faucett:2020vbu} metric to determine which EFPs from this generated set might bridge the performance gap to the PFN. ADO compares two classifiers on signal and background input pairs, measuring how often the classifiers rank the inputs in the same way. This is quantified with a Heaviside step function on many different pairs, and the results are averaged to obtain the ADO. The ADO can be interpreted as the probability that a given pair will be ordered in the same way by the two classifiers. This is intuitively similar to the AUC metric, which measures the probability that a given signal example will be ranked higher than a given background example. While AUC can be seen as comparing a classifier to the truth, the ADO compares two classifiers to one another without regard for correct ordering. To avoid training a large set of new high-level networks, one for each EFP being considered as an additional observable, we follow the strategy of Ref.~\cite{Faucett:2020vbu} and search for EFPs which have a high ADO with our PFN for the subset of events where the PFN and the high-level network disagree. In general, this process can be iterated, selecting new observables until the ADO no longer improves.

\section{Results}
\label{sec:results}

\begin{figure}
    \centering
    \includegraphics[width=0.90\linewidth]{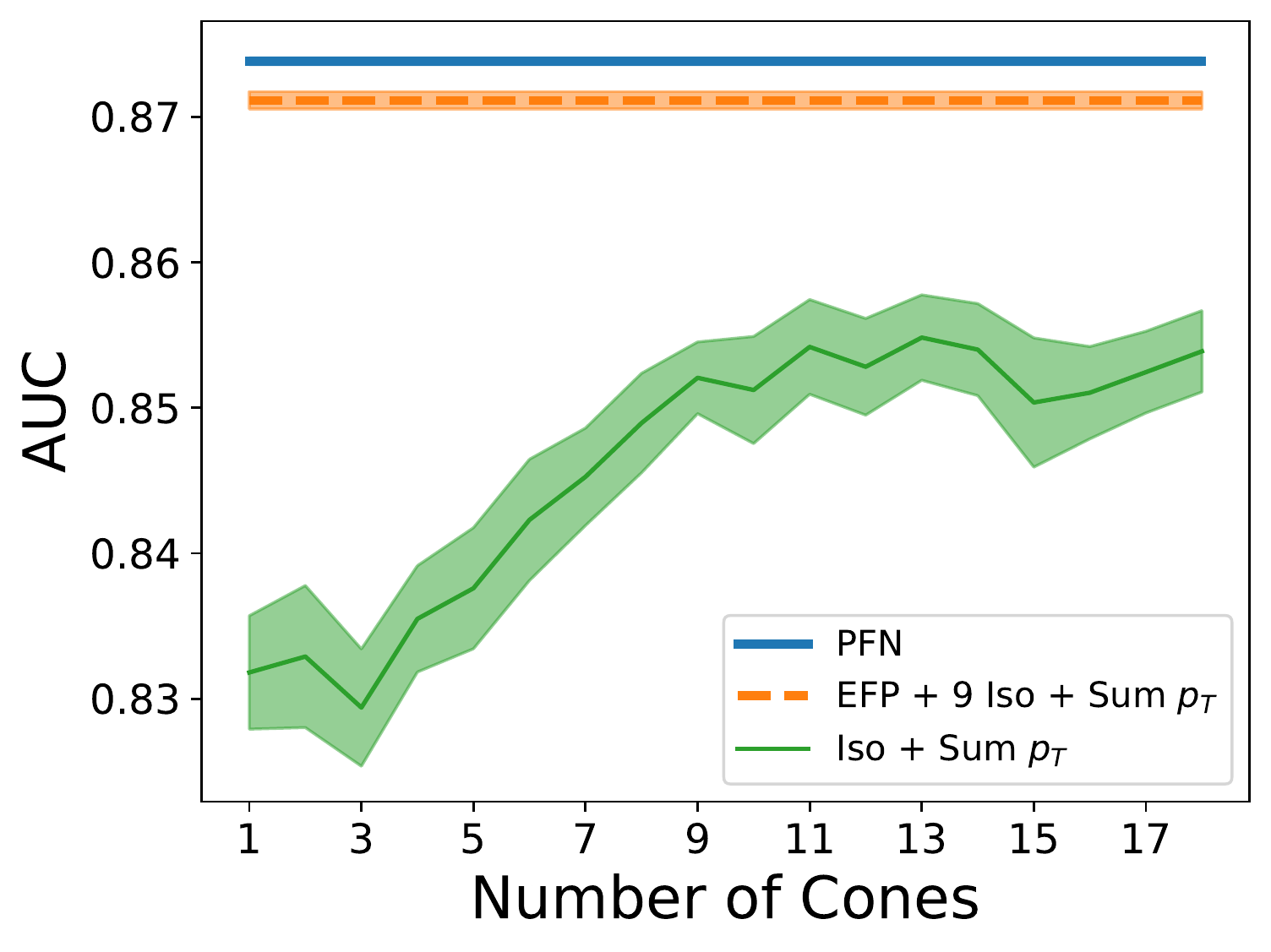}
    \includegraphics[width=0.90\linewidth]{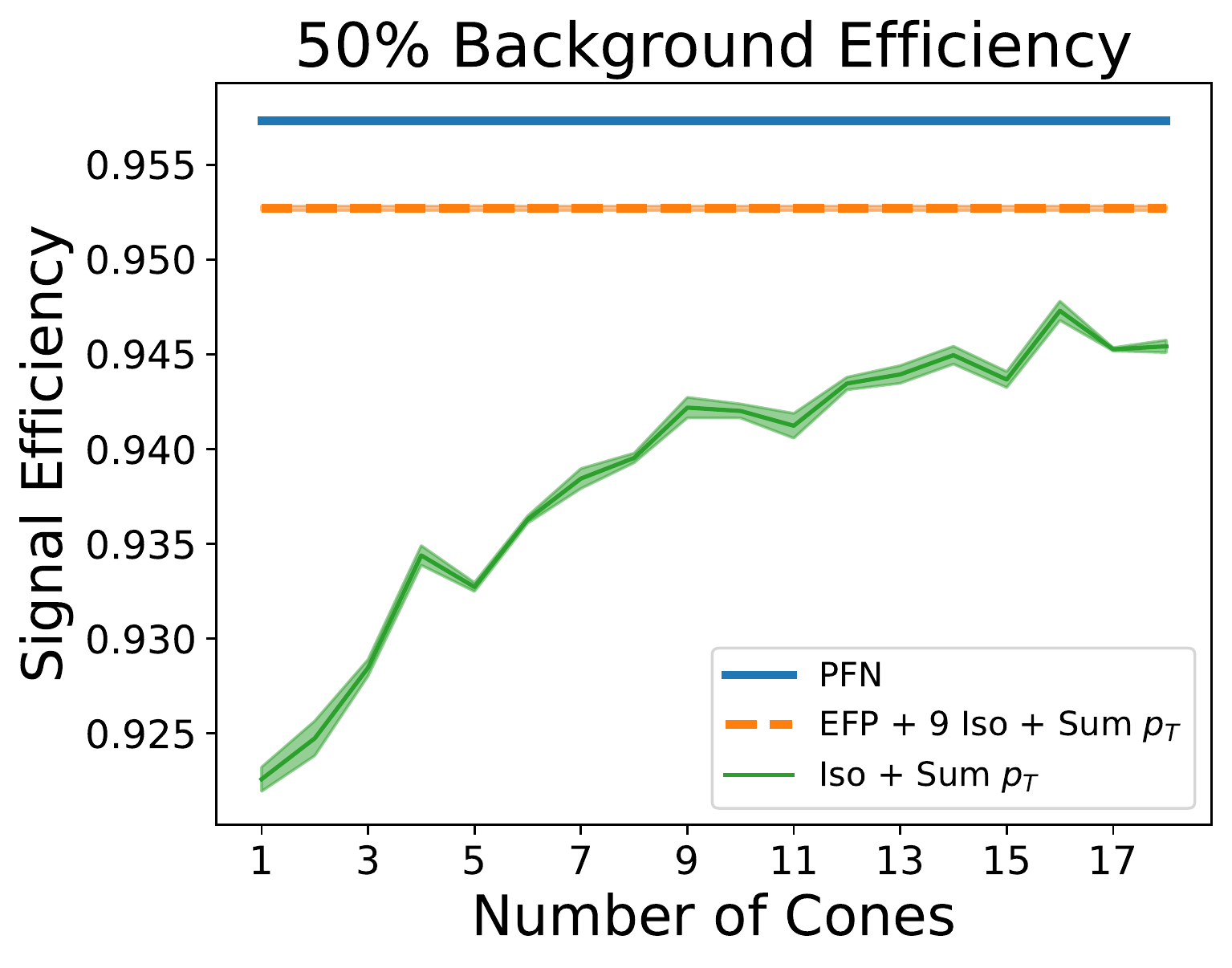}
    
    \caption{Isolation network performance shown as a function of number of input cones. Performance of the PFN and best performing high-level network are shown as benchmarks. ROC AUC is shown for each model (top) as well as the signal efficiency at a fixed background efficiency (bottom).}
    \label{fig:perf}
\end{figure}

The performance of each network is measured through ROC AUC as well as the signal efficiency at a fixed background efficiency of $50\%$. Fig.~\ref{fig:perf} illustrates the effects of including additional isolation cones as network input features. Adding cones tends to increase performance up until nine cones are used, after which there is no clear further gain in AUC. There is a significant performance gap between the network which uses nine cones and the PFN, which respectively yield AUCs of $0.848(1)$\footnote{The reported error values should be understood as rounded to $\num{1e-3}$ from values calculated to be $\lesssim\num{1e-3}$.} and $0.874(1)$, as well as signal efficiencies of $0.939(1)$ and $0.957(1)$. This suggests that isolation cones alone do not capture all discriminating information available in the low-level data. This is consistent with previous results shown on simulation~\cite{Collado:2020ehf}, and it is notable that it holds for real collider data. 

\begin{figure}
    \centering
    \begin{overpic}[width=0.90\linewidth]{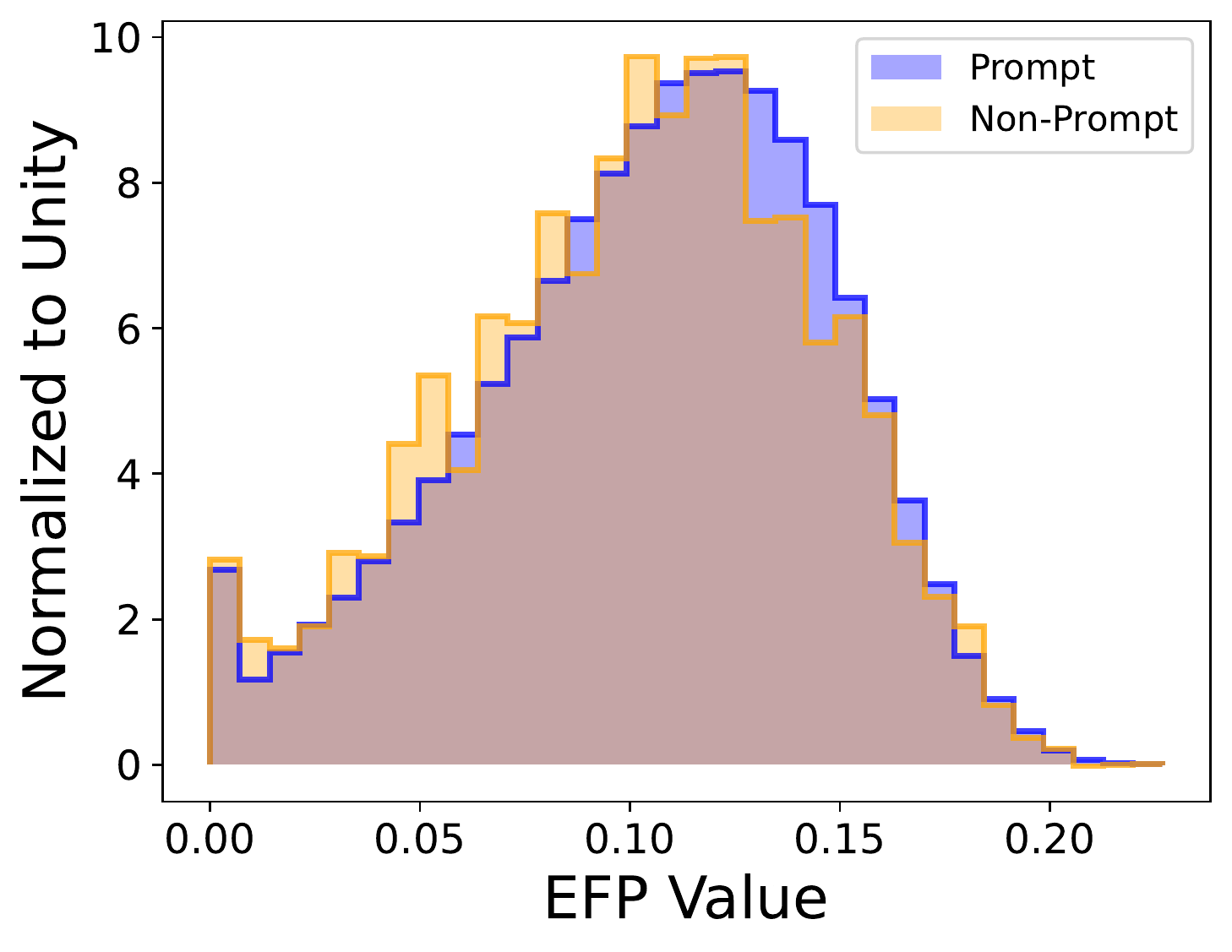}
    \put(75,35){\includegraphics[width=0.175\linewidth]{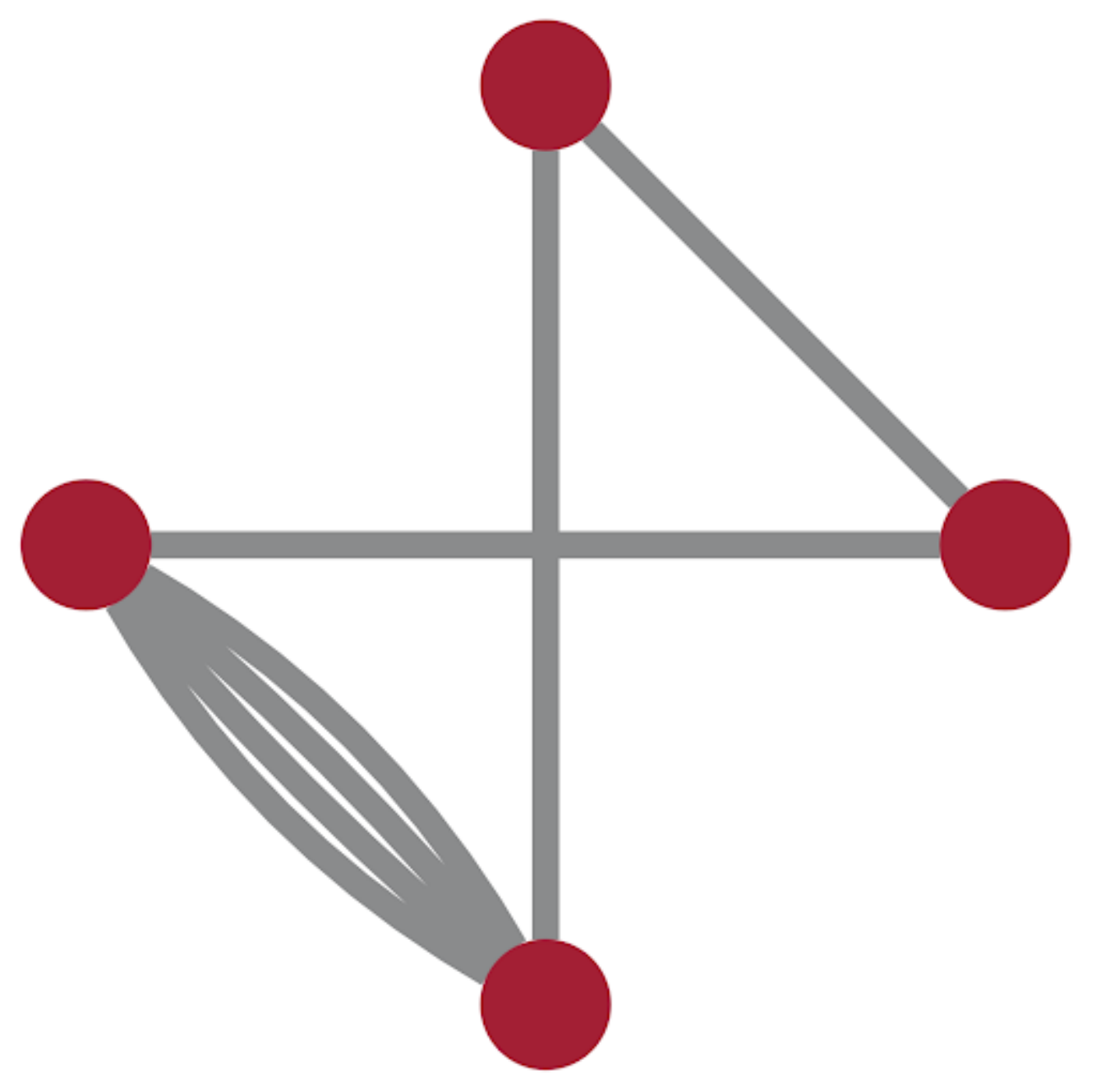}}
    \end{overpic}
    \caption{Distribution of the EFP observable identified in the search described by the text. Samples shown are separated by class using the sPlots weighting technique after applying a 50\% background efficiency cut according to the outputs of the 9 isolation cone network. Also shown is the graph representation of the EFP.} 
    \label{fig:efp_dist}
\end{figure}

We use the ADO metric to search among the EFP observables for ways to close the gap with the PFN performance. Note that the EFPs lack the built-in radial symmetry of the isolation cones, and so may contain additional useful information. The networks using EFP features are also provided the nine largest isolation cones and the summed event $p_\textrm{T}$. Remarkably, the ADO search method is able to identify a {\it single} IRC-safe EFP which obtains an AUC of $0.871(1)$ and signal efficiency of $0.953(1)$, almost fully closing the gap in AUC to the PFN from 0.026 to 0.003.
The graph representation of this EFP, as well as class distributions separated through the sPlots technique, are illustrated in Fig.~\ref{fig:efp_dist}. This EFP corresponds to parameters $\kappa=1$ and $\beta=0.25$, and the full expression is provided in Eq.~\ref{eq:selected_efp_expression}.

\begin{equation}
\sum_{a,b,c,d=1}^N
z_a z_b z_c z_d\left(
\theta_{ab}\theta_{ac}\theta_{bd}\theta_{cd}^4\right)^{1/4}
\label{eq:selected_efp_expression}
\end{equation}

An additional scan is done over the quadratic EFPs included in our full set of calculated EFPs, as these are simple in structure and are therefore more interpretable. This identifies another single EFP with $\kappa = 1$ and $\beta = 0.25$ which yields performance close to that of the one identified by the first ADO search, at an AUC of $0.870(1)$ and signal efficiency of $0.956(1)$. We further check the performance of sets of EFPs identified as useful by previous work done on simulation~\cite{Collado:2020ehf}, which selected an IRC-safe set of EFPs, as well as a set not restricted to be safe. The IRC-safe set yields an AUC of $0.868(1)$ with a signal efficiency of $0.949(1)$, while the unsafe set yields an AUC of $0.865(1)$ with a signal efficiency of $0.954(1)$. While these sets identified in simulation close much of the performance gap, they require more features and are outperformed by the EFPs identified directly on the CMS data, underscoring the importance of training in data. 

\begin{figure}
    \centering
    \includegraphics[width=0.90\linewidth]{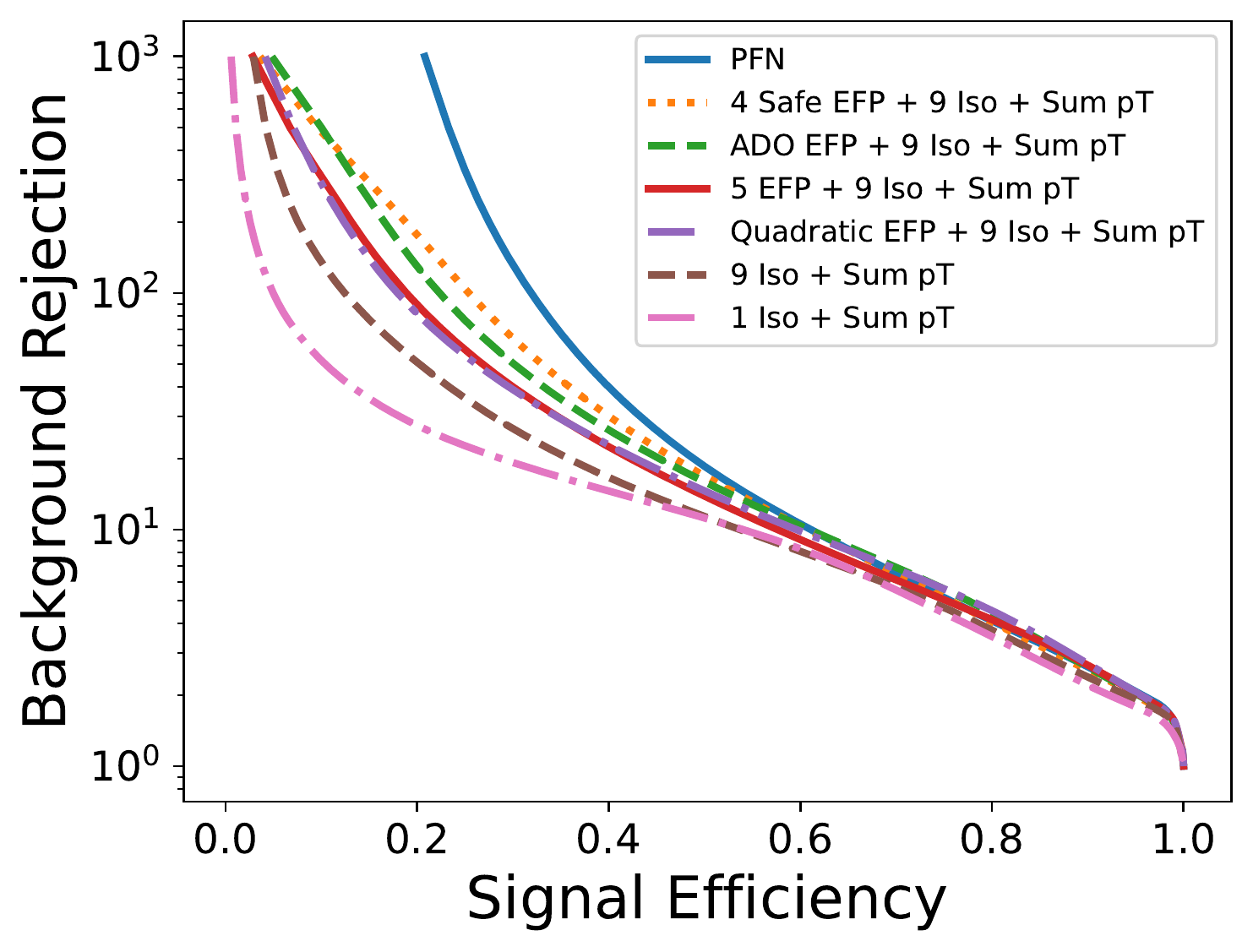}
    \caption{ Comparison of the performance of the networks described in Table~\ref{table:perf_summary}, via ROC curves.  Shown is background rejection (inverse of efficiency) versus signal efficiency.}
    \label{fig:sig_eff}
\end{figure}

A full summary of performance across the methods used is presented in Table~\ref{table:perf_summary}, as well as depicted in Fig.~\ref{fig:sig_eff}. Our results indicate that we are able to construct a minimal set of high-level observables which perform comparably to the low-level inputs, allowing for the use of more physically intuitive features and less complex networks without making concessions regarding performance.

\begin{table}[h!]
    \centering
\begin{tabular}{l|c|c|c}
\hline\hline
Input features & AUC & TPR & EFP Scan \\
\hline
   Single Iso Cone + $\sum p_{\textrm{T}}$ & $0.835$ & $0.922$ & \\
   9 Iso, $\sum p_{\textrm{T}}$ & $0.848$ & $0.939$ & \\
   9 Iso, $\sum p_{\textrm{T}}$, ADO EFP & $0.871$ & $0.953$ & CMS \\
   9 Iso, $\sum p_{\textrm{T}}$, Quadratic EFP & $0.870$ & $0.956$ & CMS \\
   9 Iso, $\sum p_{\textrm{T}}$, 4 IRC-safe EFP & $0.868$ & $0.949$ & Sim \\
   9 Iso, $\sum p_{\textrm{T}}$, 5 EFP & $0.865$ & $0.954$ &Sim \\
   Full details PFN & $0.874$ & $0.957$ & \\
   \hline\hline
\end{tabular}
    \caption{Comparison of the performance of the various networks discussed in the text. Performance is measured through ROC AUC, as well as signal efficiency (TPR) at 50\% background efficiency. Standard error is evaluated to be $\lesssim \num{1e-3}$ for both metrics over a $1\sigma$ confidence interval (see Sec.~\ref{sec:methods} for details on calculation). While the reported performance values refer only to testing done on CMS data, the ``EFP Scan'' column indicates whether the EFP inputs used were identified as useful by a scan over CMS or simulated data. These results correspond to the ROC curves in Fig~\ref{fig:sig_eff}.}
\label{table:perf_summary}
\end{table}

\section{Conclusions}
\label{sec:conclusions}

On collision data from the LHC, we apply neural networks to the problem of prompt muon discrimination. We investigate how much information is present in high and low-level representations of the data, finding that the traditionally used scalar isolation does not capture all useful classification information present at the low-level. Furthermore, we find that another high-level set of observables, the EFPs, may be used to create a network which performs almost as well as one operating at the low-level, while providing the advantage of being less complex and more human interpretable. In addition to being notable for using real rather than simulated data, this study demonstrates the use of weakly supervised training methods with CWoLa on low-level features, as well as performance evaluation without having access to individual class labels. Future work may include investigating the interpretation of the observables selected here, exploring how much information might be captured by other types of high-level observables, and the generalizability of these results.  While our study indicates that additional information is available beyond the use of simple cones, and the identification of a single EFP observable which captures that information allows for simple application and interpretation, further work would be required before implementation within an experimental context.  A robust estimate of the systematic uncertainties involved has not been done, which would be necessary to establish the optimal observables. Our result does not replace work by the experimental collaborations, but motivates further study.

\section*{Acknowledgements}

BN is supported by the U.S.~Department of Energy, Office of Science under contract DE-AC02-05CH11231. DW is supported by The Department of Energy Office of Science. The authors are grateful to Troels Petersen, David Shih, Gregor Kasieczka, Jesse Thaler, Michael Fenton and Aishik Ghosh for useful comments and suggestions.

\section*{Code and Data}

The code for this paper can be found at \href{https://github.com/Edwit4/learning_to_isolate_muons_in_data}{https://github.com/Edwit4/learning\_to\_isolate\_muons\_in\_data}.  The datasets will be provided upon reasonable request to the authors.

\bibliography{muon}

\begin{thebibliography}{37}
\expandafter\ifx\csname natexlab\endcsname\relax\def\natexlab#1{#1}\fi
\expandafter\ifx\csname bibnamefont\endcsname\relax
  \def\bibnamefont#1{#1}\fi
\expandafter\ifx\csname bibfnamefont\endcsname\relax
  \def\bibfnamefont#1{#1}\fi
\expandafter\ifx\csname citenamefont\endcsname\relax
  \def\citenamefont#1{#1}\fi
\expandafter\ifx\csname url\endcsname\relax
  \def\url#1{\texttt{#1}}\fi
\expandafter\ifx\csname urlprefix\endcsname\relax\def\urlprefix{URL }\fi
\providecommand{\bibinfo}[2]{#2}
\providecommand{\eprint}[2][]{\url{#2}}

\bibitem[{\citenamefont{Aaboud et~al.}(2018)}]{Aaboud:2017leg}
\bibinfo{author}{\bibfnamefont{M.}~\bibnamefont{Aaboud}} \bibnamefont{et~al.}
  (\bibinfo{collaboration}{ATLAS}), \bibinfo{journal}{Phys. Rev.}
  \textbf{\bibinfo{volume}{D97}}, \bibinfo{pages}{052010}
  (\bibinfo{year}{2018}), \eprint{1712.08119}.

\bibitem[{\citenamefont{Schoefbeck}(2016)}]{SCHOFBECK2016631}
\bibinfo{author}{\bibfnamefont{R.}~\bibnamefont{Schoefbeck}},
  \bibinfo{journal}{Nuclear and Particle Physics Proceedings}
  \textbf{\bibinfo{volume}{273-275}}, \bibinfo{pages}{631 }
  (\bibinfo{year}{2016}), ISSN \bibinfo{issn}{2405-6014}, \bibinfo{note}{37th
  International Conference on High Energy Physics (ICHEP)},
  \urlprefix\url{http://www.sciencedirect.com/science/article/pii/S2405601415005842}.

\bibitem[{\citenamefont{Khachatryan et~al.}(2015)}]{Khachatryan:2015kxa}
\bibinfo{author}{\bibfnamefont{V.}~\bibnamefont{Khachatryan}}
  \bibnamefont{et~al.} (\bibinfo{collaboration}{CMS}), \bibinfo{journal}{JHEP}
  \textbf{\bibinfo{volume}{11}}, \bibinfo{pages}{189} (\bibinfo{year}{2015}),
  \eprint{1508.07628}.

\bibitem[{ATL(2022)}]{ATLAS:2022hbt}
 (\bibinfo{year}{2022}), \eprint{2209.13935}.

\bibitem[{\citenamefont{Tumasyan et~al.}(2022{\natexlab{a}})}]{CMS:2021edw}
\bibinfo{author}{\bibfnamefont{A.}~\bibnamefont{Tumasyan}} \bibnamefont{et~al.}
  (\bibinfo{collaboration}{CMS}), \bibinfo{journal}{JHEP}
  \textbf{\bibinfo{volume}{04}}, \bibinfo{pages}{091}
  (\bibinfo{year}{2022}{\natexlab{a}}), \eprint{2111.06296}.

\bibitem[{\citenamefont{Aad et~al.}(2020{\natexlab{a}})}]{ATLAS:2019lng}
\bibinfo{author}{\bibfnamefont{G.}~\bibnamefont{Aad}} \bibnamefont{et~al.}
  (\bibinfo{collaboration}{ATLAS}), \bibinfo{journal}{Phys. Rev. D}
  \textbf{\bibinfo{volume}{101}}, \bibinfo{pages}{052005}
  (\bibinfo{year}{2020}{\natexlab{a}}), \eprint{1911.12606}.

\bibitem[{\citenamefont{Hoenig et~al.}(2014)\citenamefont{Hoenig, Samach, and
  Tucker-Smith}}]{Hoenig:2014}
\bibinfo{author}{\bibfnamefont{I.}~\bibnamefont{Hoenig}},
  \bibinfo{author}{\bibfnamefont{G.}~\bibnamefont{Samach}}, \bibnamefont{and}
  \bibinfo{author}{\bibfnamefont{D.}~\bibnamefont{Tucker-Smith}},
  \bibinfo{journal}{Phys. Rev. D} \textbf{\bibinfo{volume}{90}},
  \bibinfo{pages}{023} (\bibinfo{year}{2014}), \eprint{1408.1075}.

\bibitem[{ATL(2023)}]{ATLAS:2023vxg}
 (\bibinfo{year}{2023}), \eprint{2301.09342}.

\bibitem[{\citenamefont{Aaij et~al.}(2020)}]{LHCb:2020ysn}
\bibinfo{author}{\bibfnamefont{R.}~\bibnamefont{Aaij}} \bibnamefont{et~al.}
  (\bibinfo{collaboration}{LHCb}), \bibinfo{journal}{JHEP}
  \textbf{\bibinfo{volume}{10}}, \bibinfo{pages}{156} (\bibinfo{year}{2020}),
  \eprint{2007.03923}.

\bibitem[{\citenamefont{Tumasyan et~al.}(2022{\natexlab{b}})}]{CMS:2021pcy}
\bibinfo{author}{\bibfnamefont{A.}~\bibnamefont{Tumasyan}} \bibnamefont{et~al.}
  (\bibinfo{collaboration}{CMS}), \bibinfo{journal}{Eur. Phys. J. C}
  \textbf{\bibinfo{volume}{82}}, \bibinfo{pages}{290}
  (\bibinfo{year}{2022}{\natexlab{b}}), \eprint{2111.01299}.

\bibitem[{\citenamefont{Sirunyan
  et~al.}(2017{\natexlab{a}})}]{Sirunyan:2017ulk}
\bibinfo{author}{\bibfnamefont{A.~M.} \bibnamefont{Sirunyan}}
  \bibnamefont{et~al.} (\bibinfo{collaboration}{CMS}), \bibinfo{journal}{JINST}
  \textbf{\bibinfo{volume}{12}}, \bibinfo{pages}{P10003}
  (\bibinfo{year}{2017}{\natexlab{a}}), \eprint{1706.04965}.

\bibitem[{\citenamefont{Sirunyan et~al.}(2017{\natexlab{b}})}]{CMS:2017yfk}
\bibinfo{author}{\bibfnamefont{A.~M.} \bibnamefont{Sirunyan}}
  \bibnamefont{et~al.} (\bibinfo{collaboration}{CMS}), \bibinfo{journal}{JINST}
  \textbf{\bibinfo{volume}{12}}, \bibinfo{pages}{P10003}
  (\bibinfo{year}{2017}{\natexlab{b}}), \eprint{1706.04965}.

\bibitem[{\citenamefont{Aad et~al.}(2021)}]{ATLAS:2020auj}
\bibinfo{author}{\bibfnamefont{G.}~\bibnamefont{Aad}} \bibnamefont{et~al.}
  (\bibinfo{collaboration}{ATLAS}), \bibinfo{journal}{Eur. Phys. J. C}
  \textbf{\bibinfo{volume}{81}}, \bibinfo{pages}{578} (\bibinfo{year}{2021}),
  \eprint{2012.00578}.

\bibitem[{\citenamefont{Baldi et~al.}(2014)\citenamefont{Baldi, Sadowski, and
  Whiteson}}]{Baldi:2014kfa}
\bibinfo{author}{\bibfnamefont{P.}~\bibnamefont{Baldi}},
  \bibinfo{author}{\bibfnamefont{P.}~\bibnamefont{Sadowski}}, \bibnamefont{and}
  \bibinfo{author}{\bibfnamefont{D.}~\bibnamefont{Whiteson}},
  \bibinfo{journal}{Nature Commun.} \textbf{\bibinfo{volume}{5}},
  \bibinfo{pages}{4308} (\bibinfo{year}{2014}), \eprint{1402.4735}.

\bibitem[{\citenamefont{de~Oliveira et~al.}(2016)\citenamefont{de~Oliveira,
  Kagan, Mackey, Nachman, and Schwartzman}}]{deOliveira:2015xxd}
\bibinfo{author}{\bibfnamefont{L.}~\bibnamefont{de~Oliveira}},
  \bibinfo{author}{\bibfnamefont{M.}~\bibnamefont{Kagan}},
  \bibinfo{author}{\bibfnamefont{L.}~\bibnamefont{Mackey}},
  \bibinfo{author}{\bibfnamefont{B.}~\bibnamefont{Nachman}}, \bibnamefont{and}
  \bibinfo{author}{\bibfnamefont{A.}~\bibnamefont{Schwartzman}},
  \bibinfo{journal}{JHEP} \textbf{\bibinfo{volume}{07}}, \bibinfo{pages}{069}
  (\bibinfo{year}{2016}), \eprint{1511.05190}.

\bibitem[{\citenamefont{Feickert and Nachman}(2021)}]{Feickert:2021ajf}
\bibinfo{author}{\bibfnamefont{M.}~\bibnamefont{Feickert}} \bibnamefont{and}
  \bibinfo{author}{\bibfnamefont{B.}~\bibnamefont{Nachman}}
  (\bibinfo{year}{2021}), \eprint{2102.02770}.

\bibitem[{\citenamefont{Collado et~al.}(2020)\citenamefont{Collado, Bauer,
  Witkowski, Faucett, Whiteson, and Baldi}}]{Collado:2020ehf}
\bibinfo{author}{\bibfnamefont{J.}~\bibnamefont{Collado}},
  \bibinfo{author}{\bibfnamefont{K.}~\bibnamefont{Bauer}},
  \bibinfo{author}{\bibfnamefont{E.}~\bibnamefont{Witkowski}},
  \bibinfo{author}{\bibfnamefont{T.}~\bibnamefont{Faucett}},
  \bibinfo{author}{\bibfnamefont{D.}~\bibnamefont{Whiteson}}, \bibnamefont{and}
  \bibinfo{author}{\bibfnamefont{P.}~\bibnamefont{Baldi}},
  \bibinfo{journal}{JHEP} \textbf{\bibinfo{volume}{21}}, \bibinfo{pages}{200}
  (\bibinfo{year}{2020}), \eprint{2102.02278}.

\bibitem[{\citenamefont{Aad et~al.}(2016)}]{ATLAS:2016lqx}
\bibinfo{author}{\bibfnamefont{G.}~\bibnamefont{Aad}} \bibnamefont{et~al.}
  (\bibinfo{collaboration}{ATLAS}), \bibinfo{journal}{Eur. Phys. J. C}
  \textbf{\bibinfo{volume}{76}}, \bibinfo{pages}{292} (\bibinfo{year}{2016}),
  \eprint{1603.05598}.

\bibitem[{\citenamefont{Sirunyan et~al.}(2018)}]{CMS:2018rym}
\bibinfo{author}{\bibfnamefont{A.~M.} \bibnamefont{Sirunyan}}
  \bibnamefont{et~al.} (\bibinfo{collaboration}{CMS}), \bibinfo{journal}{JINST}
  \textbf{\bibinfo{volume}{13}}, \bibinfo{pages}{P06015}
  (\bibinfo{year}{2018}), \eprint{1804.04528}.

\bibitem[{\citenamefont{Cesarotti et~al.}(2019)\citenamefont{Cesarotti, Soreq,
  Strassler, Thaler, and Xue}}]{Cesarotti:2019nax}
\bibinfo{author}{\bibfnamefont{C.}~\bibnamefont{Cesarotti}},
  \bibinfo{author}{\bibfnamefont{Y.}~\bibnamefont{Soreq}},
  \bibinfo{author}{\bibfnamefont{M.~J.} \bibnamefont{Strassler}},
  \bibinfo{author}{\bibfnamefont{J.}~\bibnamefont{Thaler}}, \bibnamefont{and}
  \bibinfo{author}{\bibfnamefont{W.}~\bibnamefont{Xue}},
  \bibinfo{journal}{Phys. Rev. D} \textbf{\bibinfo{volume}{100}},
  \bibinfo{pages}{015021} (\bibinfo{year}{2019}), \eprint{1902.04222}.

\bibitem[{\citenamefont{Metodiev et~al.}(2017)\citenamefont{Metodiev, Nachman,
  and Thaler}}]{Metodiev:2017vrx}
\bibinfo{author}{\bibfnamefont{E.~M.} \bibnamefont{Metodiev}},
  \bibinfo{author}{\bibfnamefont{B.}~\bibnamefont{Nachman}}, \bibnamefont{and}
  \bibinfo{author}{\bibfnamefont{J.}~\bibnamefont{Thaler}},
  \bibinfo{journal}{JHEP} \textbf{\bibinfo{volume}{10}}, \bibinfo{pages}{174}
  (\bibinfo{year}{2017}), \eprint{1708.02949}.

\bibitem[{\citenamefont{Sirunyan et~al.}(2020)}]{CMS:2019eih}
\bibinfo{author}{\bibfnamefont{A.~M.} \bibnamefont{Sirunyan}}
  \bibnamefont{et~al.} (\bibinfo{collaboration}{CMS}), \bibinfo{journal}{Phys.
  Lett. B} \textbf{\bibinfo{volume}{803}}, \bibinfo{pages}{135285}
  (\bibinfo{year}{2020}), \eprint{1909.05306}.

\bibitem[{\citenamefont{Mikuni}(2021)}]{Mikuni:2781479}
\bibinfo{author}{\bibfnamefont{V.~M.} \bibnamefont{Mikuni}},
  \emph{\bibinfo{title}{{Collider Physics Measurements in High Jet Multiplicity
  Final States}}} (\bibinfo{year}{2021}), \bibinfo{note}{presented 2021},
  \urlprefix\url{https://cds.cern.ch/record/2781479}.

\bibitem[{\citenamefont{Collins et~al.}(2019)\citenamefont{Collins, Howe, and
  Nachman}}]{Collins_2019}
\bibinfo{author}{\bibfnamefont{J.~H.} \bibnamefont{Collins}},
  \bibinfo{author}{\bibfnamefont{K.}~\bibnamefont{Howe}}, \bibnamefont{and}
  \bibinfo{author}{\bibfnamefont{B.}~\bibnamefont{Nachman}},
  \bibinfo{journal}{Physical Review D} \textbf{\bibinfo{volume}{99}}
  (\bibinfo{year}{2019}),
  \urlprefix\url{https://doi.org/10.1103%2Fphysrevd.99.014038}.

\bibitem[{\citenamefont{Collins et~al.}(2018)\citenamefont{Collins, Howe, and
  Nachman}}]{Collins:2018epr}
\bibinfo{author}{\bibfnamefont{J.~H.} \bibnamefont{Collins}},
  \bibinfo{author}{\bibfnamefont{K.}~\bibnamefont{Howe}}, \bibnamefont{and}
  \bibinfo{author}{\bibfnamefont{B.}~\bibnamefont{Nachman}},
  \bibinfo{journal}{Phys. Rev. Lett.} \textbf{\bibinfo{volume}{121}},
  \bibinfo{pages}{241803} (\bibinfo{year}{2018}), \eprint{1805.02664}.

\bibitem[{\citenamefont{Aad et~al.}(2020{\natexlab{b}})}]{ATLAS:2020iwa}
\bibinfo{author}{\bibfnamefont{G.}~\bibnamefont{Aad}} \bibnamefont{et~al.}
  (\bibinfo{collaboration}{ATLAS}), \bibinfo{journal}{Phys. Rev. Lett.}
  \textbf{\bibinfo{volume}{125}}, \bibinfo{pages}{131801}
  (\bibinfo{year}{2020}{\natexlab{b}}), \eprint{2005.02983}.

\bibitem[{\citenamefont{Komiske et~al.}(2018)\citenamefont{Komiske, Metodiev,
  and Thaler}}]{Komiske:2017aww}
\bibinfo{author}{\bibfnamefont{P.~T.} \bibnamefont{Komiske}},
  \bibinfo{author}{\bibfnamefont{E.~M.} \bibnamefont{Metodiev}},
  \bibnamefont{and} \bibinfo{author}{\bibfnamefont{J.}~\bibnamefont{Thaler}},
  \bibinfo{journal}{JHEP} \textbf{\bibinfo{volume}{04}}, \bibinfo{pages}{013}
  (\bibinfo{year}{2018}), \eprint{1712.07124}.

\bibitem[{\citenamefont{Faucett et~al.}(2020)\citenamefont{Faucett, Thaler, and
  Whiteson}}]{Faucett:2020vbu}
\bibinfo{author}{\bibfnamefont{T.}~\bibnamefont{Faucett}},
  \bibinfo{author}{\bibfnamefont{J.}~\bibnamefont{Thaler}}, \bibnamefont{and}
  \bibinfo{author}{\bibfnamefont{D.}~\bibnamefont{Whiteson}}
  (\bibinfo{year}{2020}), \eprint{2010.11998}.

\bibitem[{\citenamefont{Chatrchyan et~al.}(2008)}]{CMS:2008xjf}
\bibinfo{author}{\bibfnamefont{S.}~\bibnamefont{Chatrchyan}}
  \bibnamefont{et~al.} (\bibinfo{collaboration}{CMS}), \bibinfo{journal}{JINST}
  \textbf{\bibinfo{volume}{3}}, \bibinfo{pages}{S08004} (\bibinfo{year}{2008}).

\bibitem[{cms()}]{cms-open-data}
\emph{\bibinfo{title}{{CERN Open Data Portal}}},
  \urlprefix\url{http://opendata.cern.ch}.

\bibitem[{\citenamefont{Workman et~al.}(2022)}]{ParticleDataGroup:2022pth}
\bibinfo{author}{\bibfnamefont{R.~L.} \bibnamefont{Workman}}
  \bibnamefont{et~al.} (\bibinfo{collaboration}{Particle Data Group}),
  \bibinfo{journal}{PTEP} \textbf{\bibinfo{volume}{2022}},
  \bibinfo{pages}{083C01} (\bibinfo{year}{2022}).

\bibitem[{\citenamefont{Virtanen et~al.}(2020)\citenamefont{Virtanen, Gommers,
  Oliphant, Haberland, Reddy, Cournapeau, Burovski, Peterson, Weckesser, Bright
  et~al.}}]{2020SciPy-NMeth}
\bibinfo{author}{\bibfnamefont{P.}~\bibnamefont{Virtanen}},
  \bibinfo{author}{\bibfnamefont{R.}~\bibnamefont{Gommers}},
  \bibinfo{author}{\bibfnamefont{T.~E.} \bibnamefont{Oliphant}},
  \bibinfo{author}{\bibfnamefont{M.}~\bibnamefont{Haberland}},
  \bibinfo{author}{\bibfnamefont{T.}~\bibnamefont{Reddy}},
  \bibinfo{author}{\bibfnamefont{D.}~\bibnamefont{Cournapeau}},
  \bibinfo{author}{\bibfnamefont{E.}~\bibnamefont{Burovski}},
  \bibinfo{author}{\bibfnamefont{P.}~\bibnamefont{Peterson}},
  \bibinfo{author}{\bibfnamefont{W.}~\bibnamefont{Weckesser}},
  \bibinfo{author}{\bibfnamefont{J.}~\bibnamefont{Bright}},
  \bibnamefont{et~al.}, \bibinfo{journal}{Nature Methods}
  \textbf{\bibinfo{volume}{17}}, \bibinfo{pages}{261} (\bibinfo{year}{2020}).

\bibitem[{\citenamefont{Pivk and Le~Diberder}(2005)}]{Pivk:2004ty}
\bibinfo{author}{\bibfnamefont{M.}~\bibnamefont{Pivk}} \bibnamefont{and}
  \bibinfo{author}{\bibfnamefont{F.~R.} \bibnamefont{Le~Diberder}},
  \bibinfo{journal}{Nucl. Instrum. Meth. A} \textbf{\bibinfo{volume}{555}},
  \bibinfo{pages}{356} (\bibinfo{year}{2005}), \eprint{physics/0402083}.

\bibitem[{\citenamefont{Nachman}(2020)}]{Nachman:2019dol}
\bibinfo{author}{\bibfnamefont{B.}~\bibnamefont{Nachman}},
  \bibinfo{journal}{SciPost Phys.} \textbf{\bibinfo{volume}{8}},
  \bibinfo{pages}{090} (\bibinfo{year}{2020}), \eprint{1909.03081}.

\bibitem[{\citenamefont{Efron}(1979)}]{10.1214/aos/1176344552}
\bibinfo{author}{\bibfnamefont{B.}~\bibnamefont{Efron}}, \bibinfo{journal}{The
  Annals of Statistics} \textbf{\bibinfo{volume}{7}}, \bibinfo{pages}{1 }
  (\bibinfo{year}{1979}),
  \urlprefix\url{https://doi.org/10.1214/aos/1176344552}.

\bibitem[{\citenamefont{Zaheer et~al.}(2018)\citenamefont{Zaheer, Kottur,
  Ravanbakhsh, Poczos, Salakhutdinov, and Smola}}]{zaheer2018deep}
\bibinfo{author}{\bibfnamefont{M.}~\bibnamefont{Zaheer}},
  \bibinfo{author}{\bibfnamefont{S.}~\bibnamefont{Kottur}},
  \bibinfo{author}{\bibfnamefont{S.}~\bibnamefont{Ravanbakhsh}},
  \bibinfo{author}{\bibfnamefont{B.}~\bibnamefont{Poczos}},
  \bibinfo{author}{\bibfnamefont{R.}~\bibnamefont{Salakhutdinov}},
  \bibnamefont{and} \bibinfo{author}{\bibfnamefont{A.}~\bibnamefont{Smola}},
  \emph{\bibinfo{title}{Deep sets}} (\bibinfo{year}{2018}),
  \eprint{1703.06114}.

\bibitem[{\citenamefont{Komiske et~al.}(2019)\citenamefont{Komiske, Metodiev,
  and Thaler}}]{Komiske:2018cqr}
\bibinfo{author}{\bibfnamefont{P.~T.} \bibnamefont{Komiske}},
  \bibinfo{author}{\bibfnamefont{E.~M.} \bibnamefont{Metodiev}},
  \bibnamefont{and} \bibinfo{author}{\bibfnamefont{J.}~\bibnamefont{Thaler}},
  \bibinfo{journal}{JHEP} \textbf{\bibinfo{volume}{01}}, \bibinfo{pages}{121}
  (\bibinfo{year}{2019}), \eprint{1810.05165}.

\end{thebibliography}

\appendix
\section{APPENDIX}
\label{sec:app}

 CWoLa assumes that the mixed samples are generated in such a way that a given component feature is distributed the same way in one sample as it is in the other. While we cannot explicitly demonstrate this on an unlabeled dataset, we can use a simulated dataset similar to the experimental data to probe whether we can reasonably expect this assumption to hold. 
 
\begin{figure}
\includegraphics[width=0.90\linewidth]{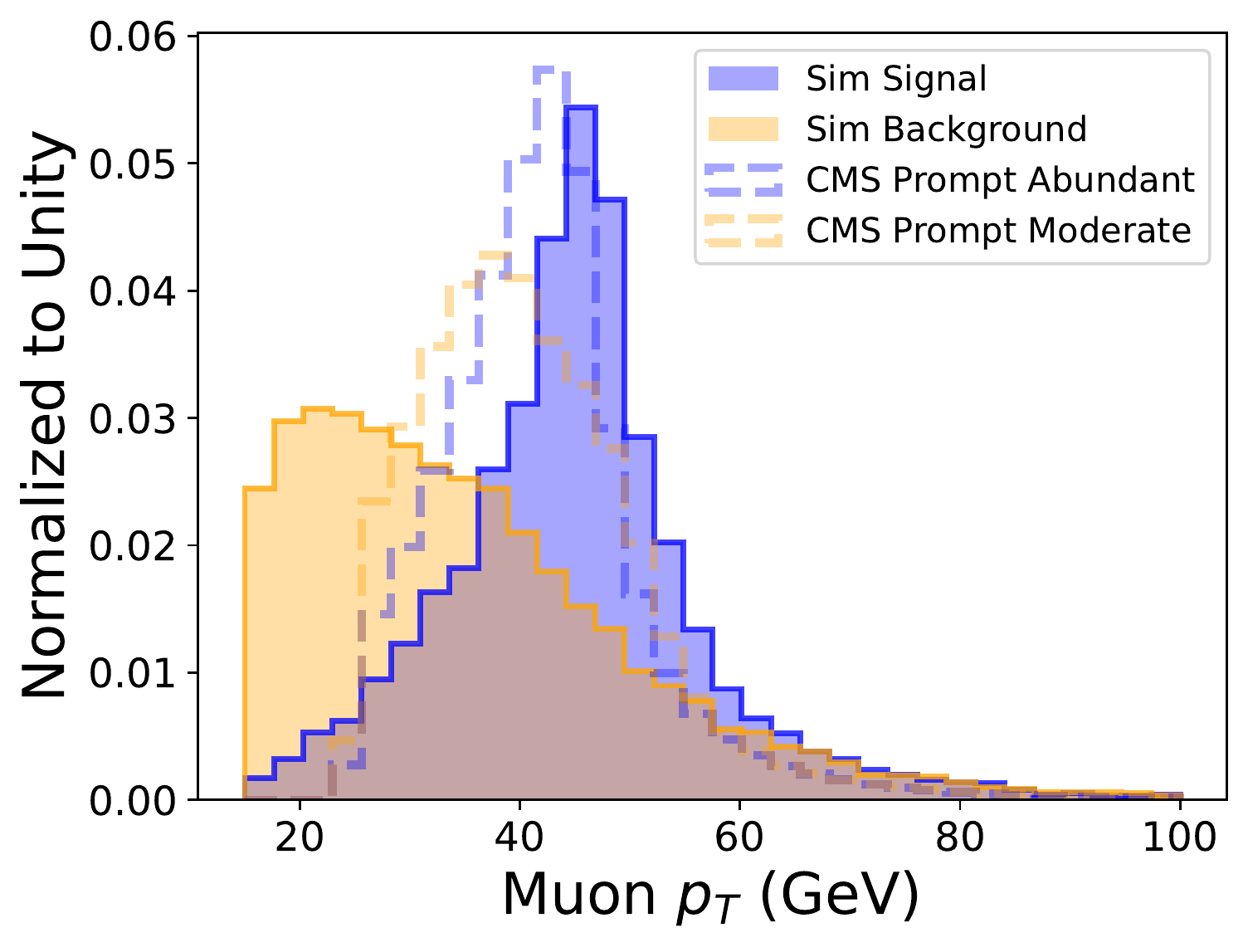}
\includegraphics[width=0.90\linewidth]{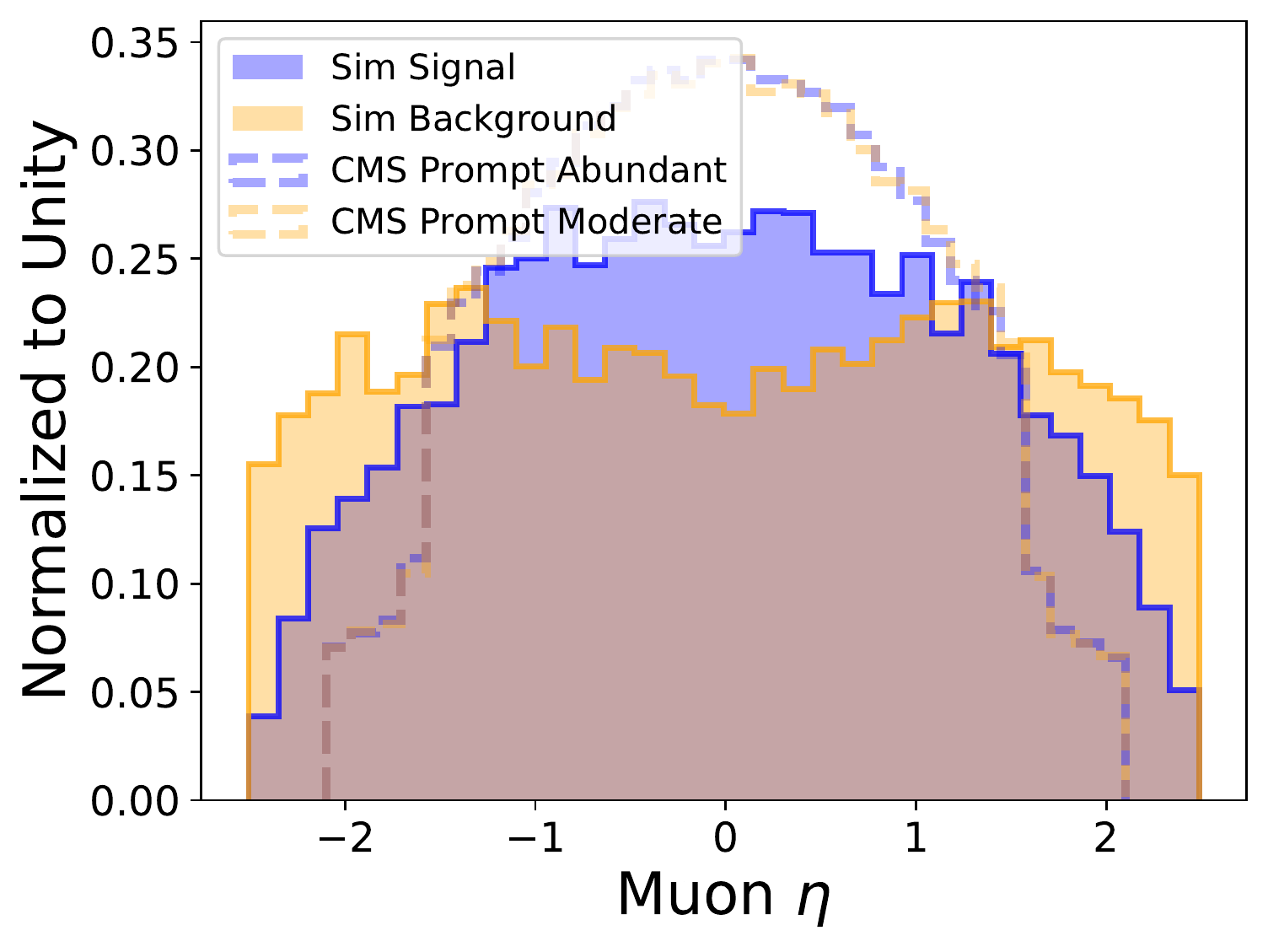}
\caption {(MG5+Pythia+Delphes) Distributions of the simulated muon transverse momentum and pseudorapidity.
}
\label{fig:sim_muon_dists}
\end{figure}

\begin{figure}
\includegraphics[width=0.95\linewidth]{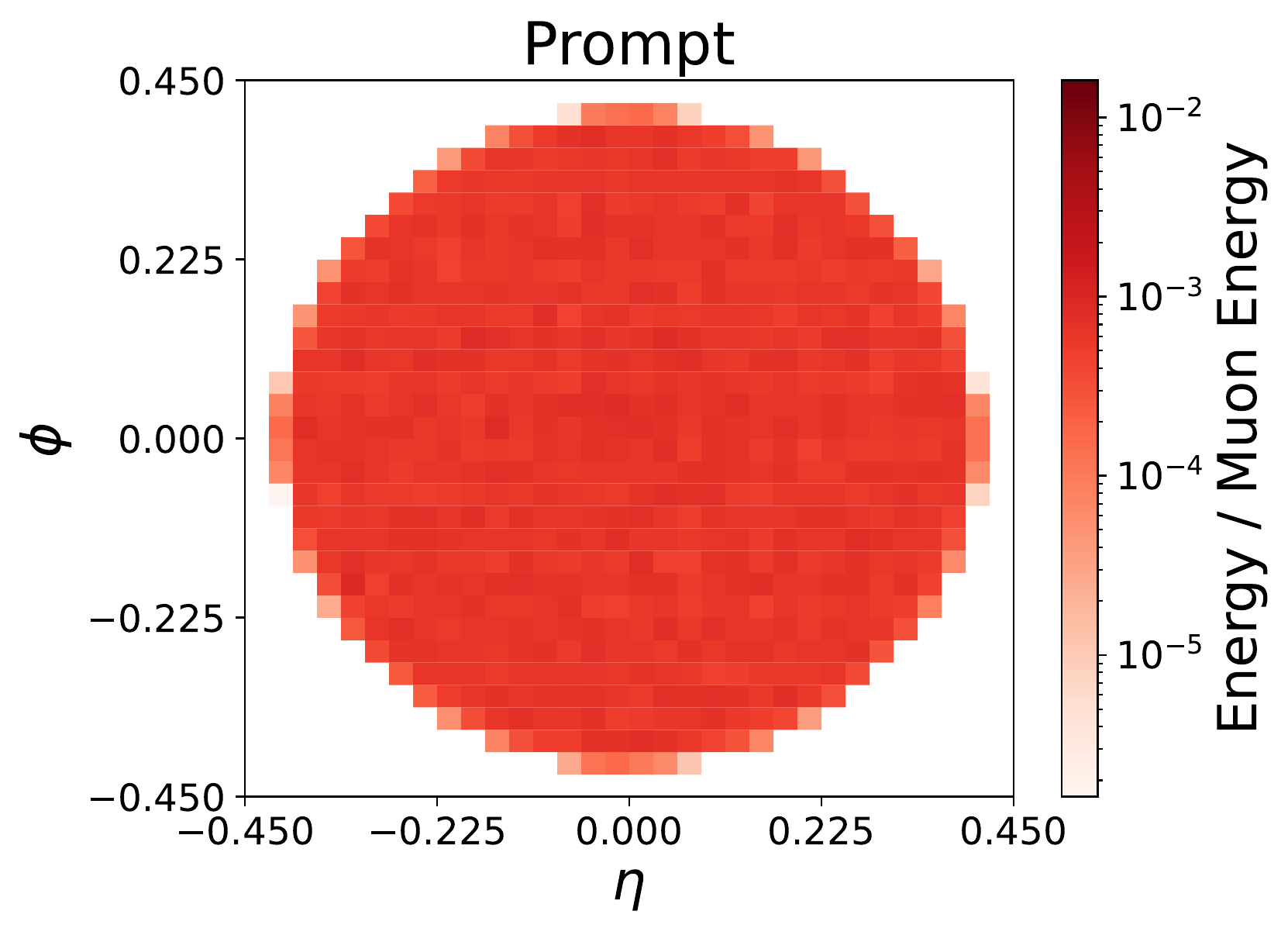}
\includegraphics[width=0.95\linewidth]{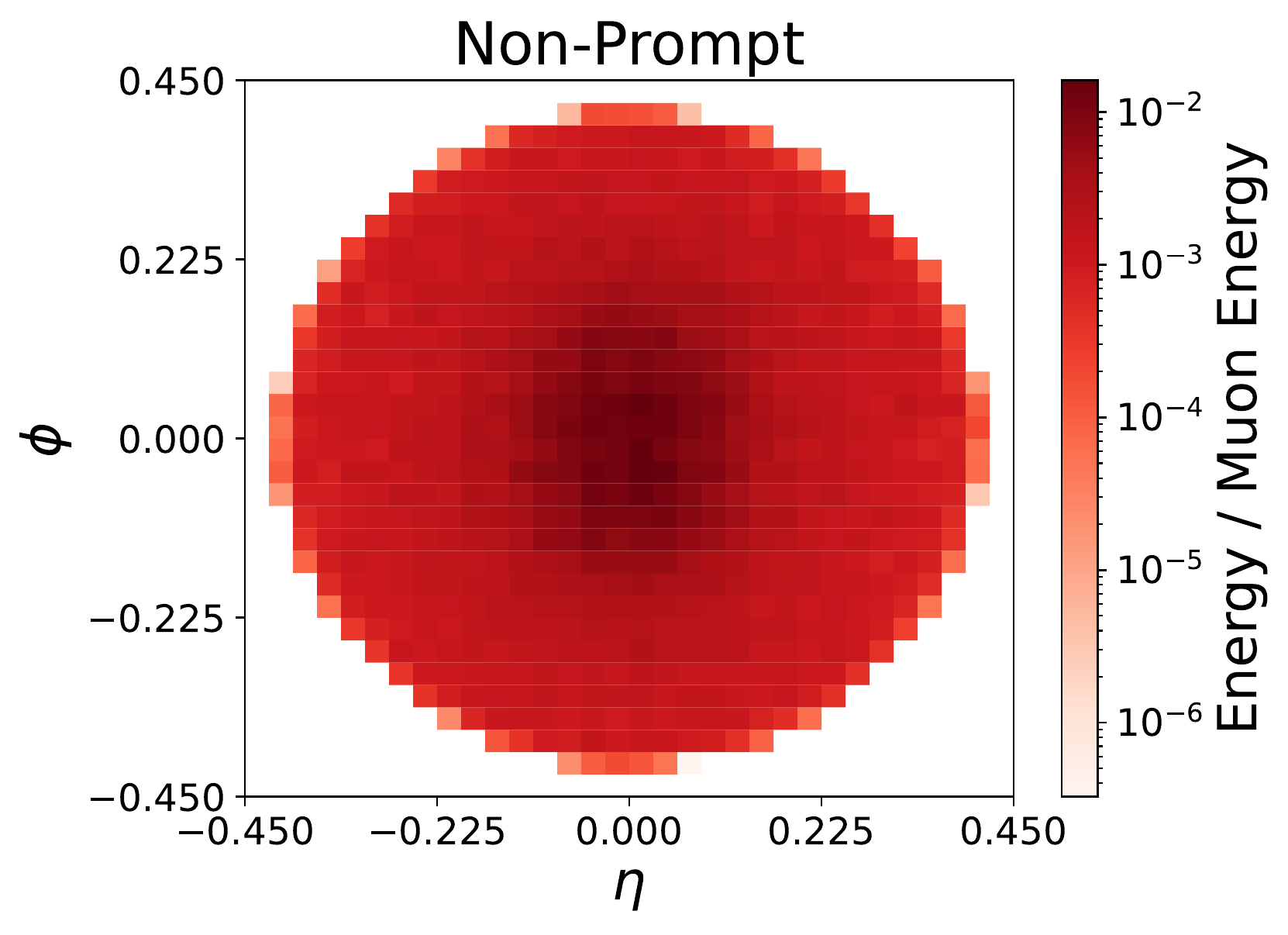}
\caption {(MG5+Pythia+Delphes) Average event images similar to Fig~\ref{fig:average_images}, but for the simulated dataset and separated by prompt and non-prompt events. 
}
\label{fig:sim_avg_images}
\end{figure}

 We simulate events where prompt muons are generated by the process $pp \rightarrow Z \rightarrow \mu^+\mu^-$, and non-prompt muons by $pp \rightarrow b\bar{b}$. A center of mass energy of $s=\sqrt(13)$ TeV is used. Madgraph5, Pythia, and Delphes are used respectively for collision and heavy boson decay simulation, showering and hadronization, and the detector simulation, with pile-up included. In total we generate 22766 events, where half are prompt and the other half are non-prompt events. The muon transverse momentum and pseudorapidity distributions for this dataset are shown in Fig~\ref{fig:sim_muon_dists}, and the average event images are shown in Fig~\ref{fig:sim_avg_images}, where quantities are separated between the prompt and non-prompt distributions. 
 
\begin{figure}
\includegraphics[width=0.90\linewidth]{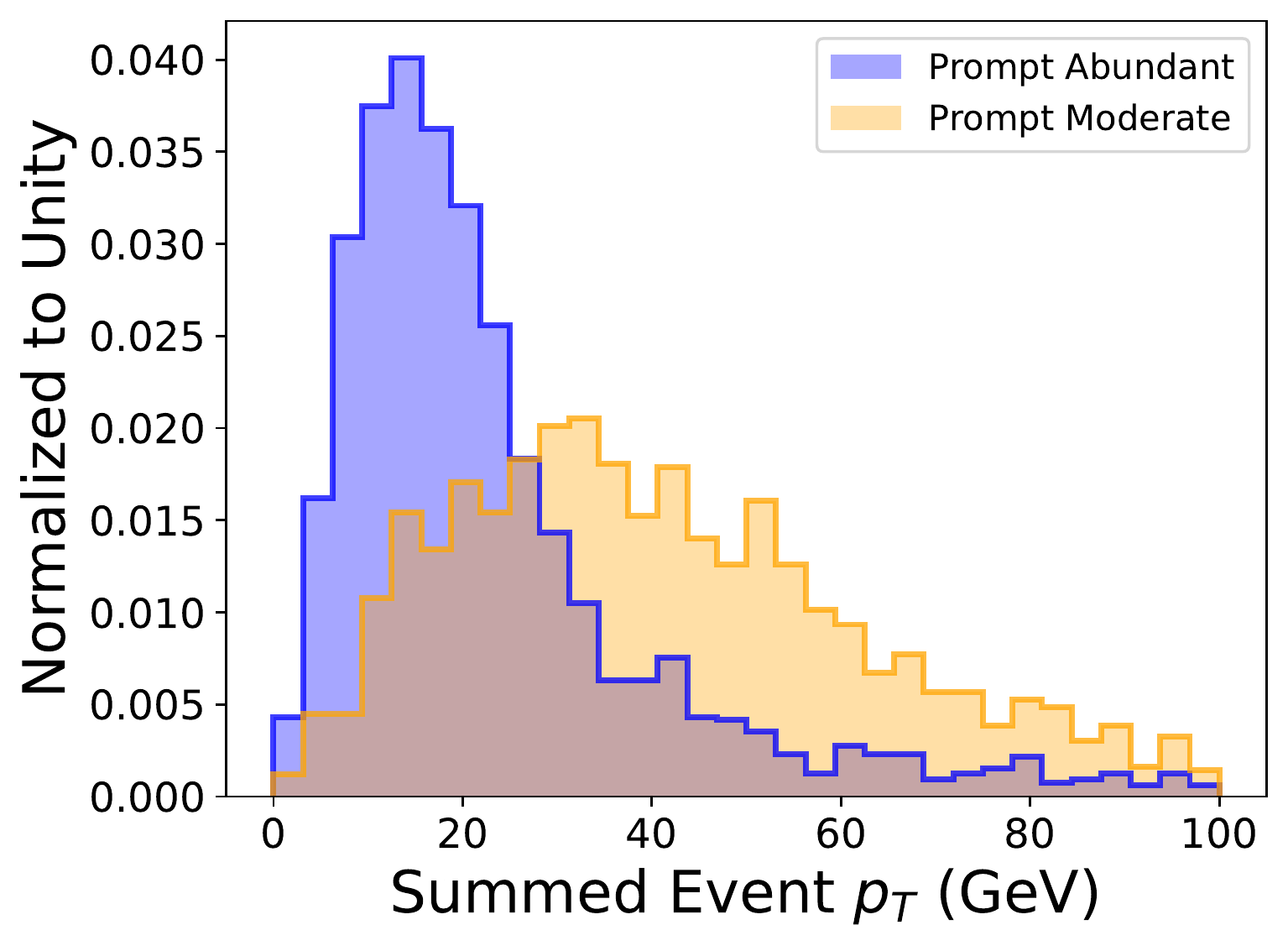}
\includegraphics[width=0.90\linewidth]{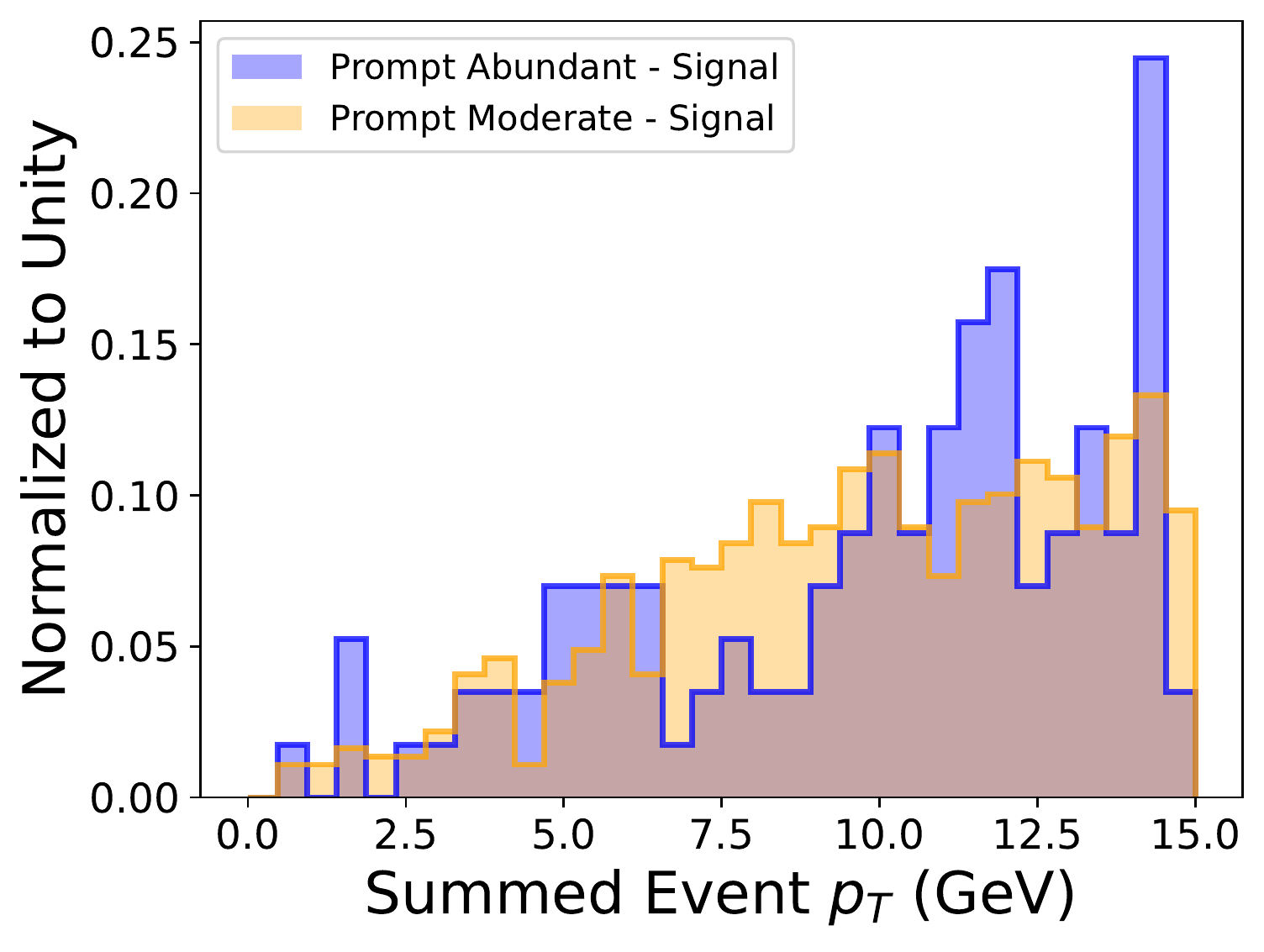}
\includegraphics[width=0.90\linewidth]{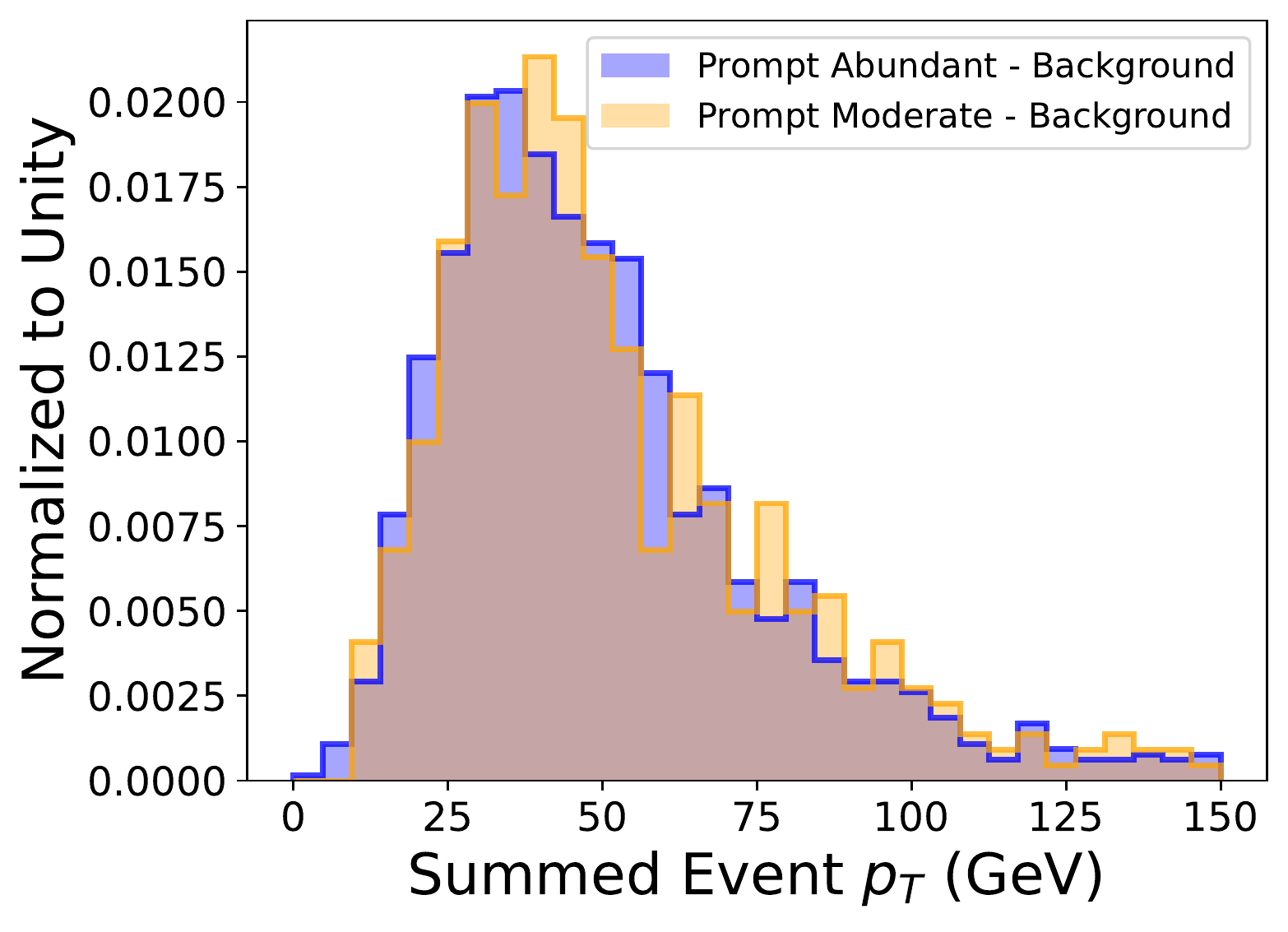}
\caption {(MG5+Pythia+Delphes) (Top) The total summed event $p_{\mathrm{T}}$ distributions for two simulated mixed samples. (Middle) Only the signal components of the two simulated mixed samples.
(Bottom) Only the background components of the two simulated mixed samples.
We see that while the class proportions are different, the signal and background distributions are approximately the same across the samples.
}
\label{fig:sim_pT_mixed_samples}
\end{figure}

\begin{figure}
\includegraphics[width=0.90\linewidth]{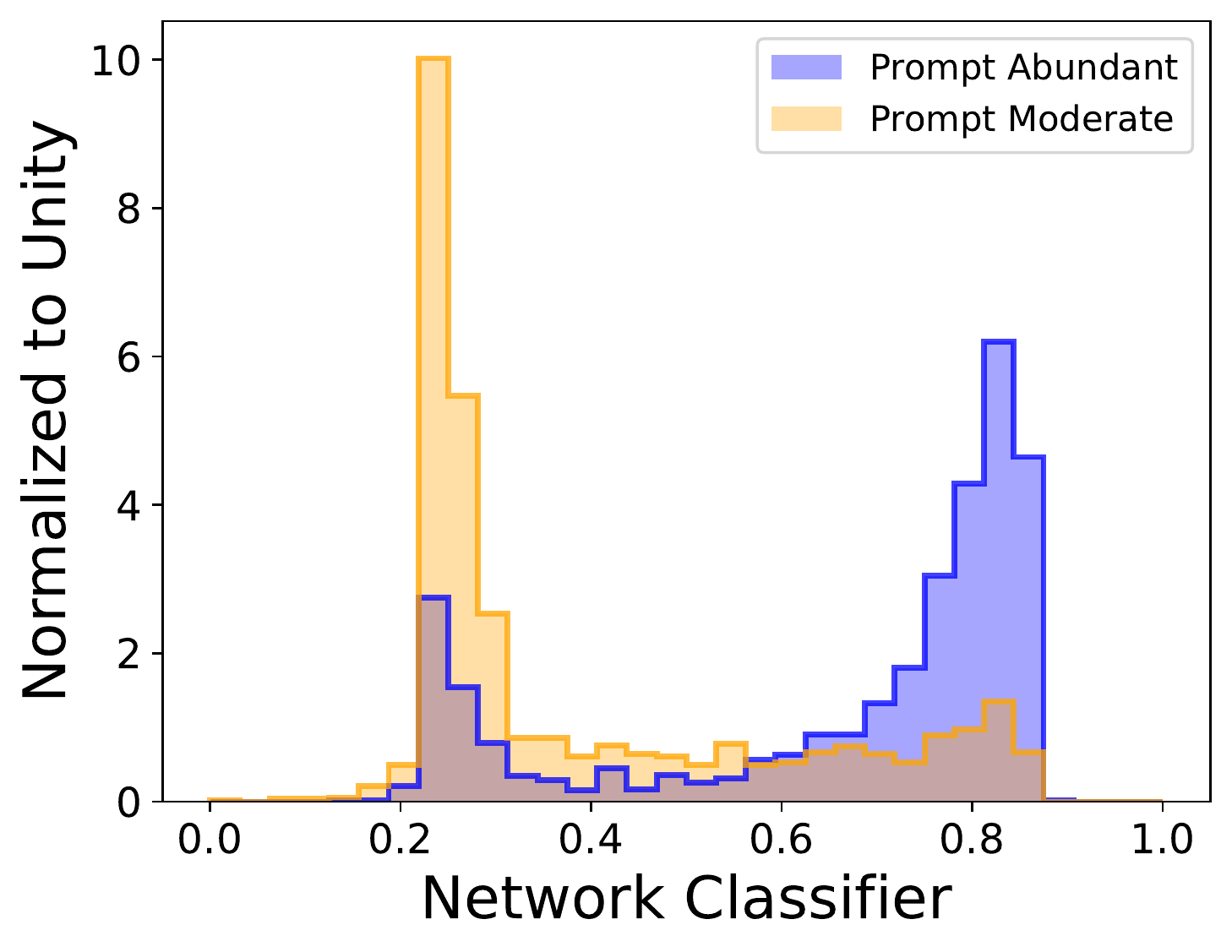}
\includegraphics[width=0.90\linewidth]{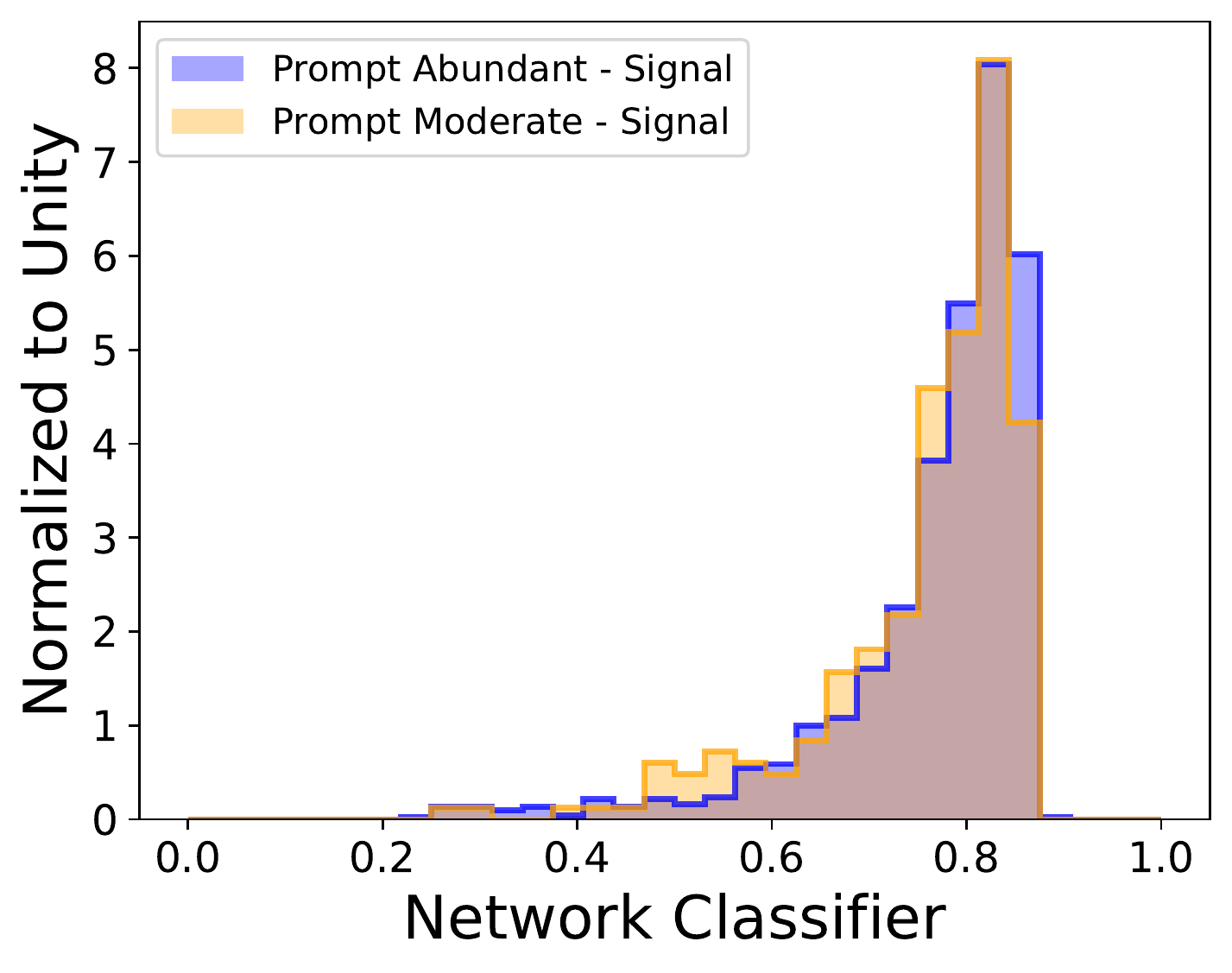}
\includegraphics[width=0.90\linewidth]{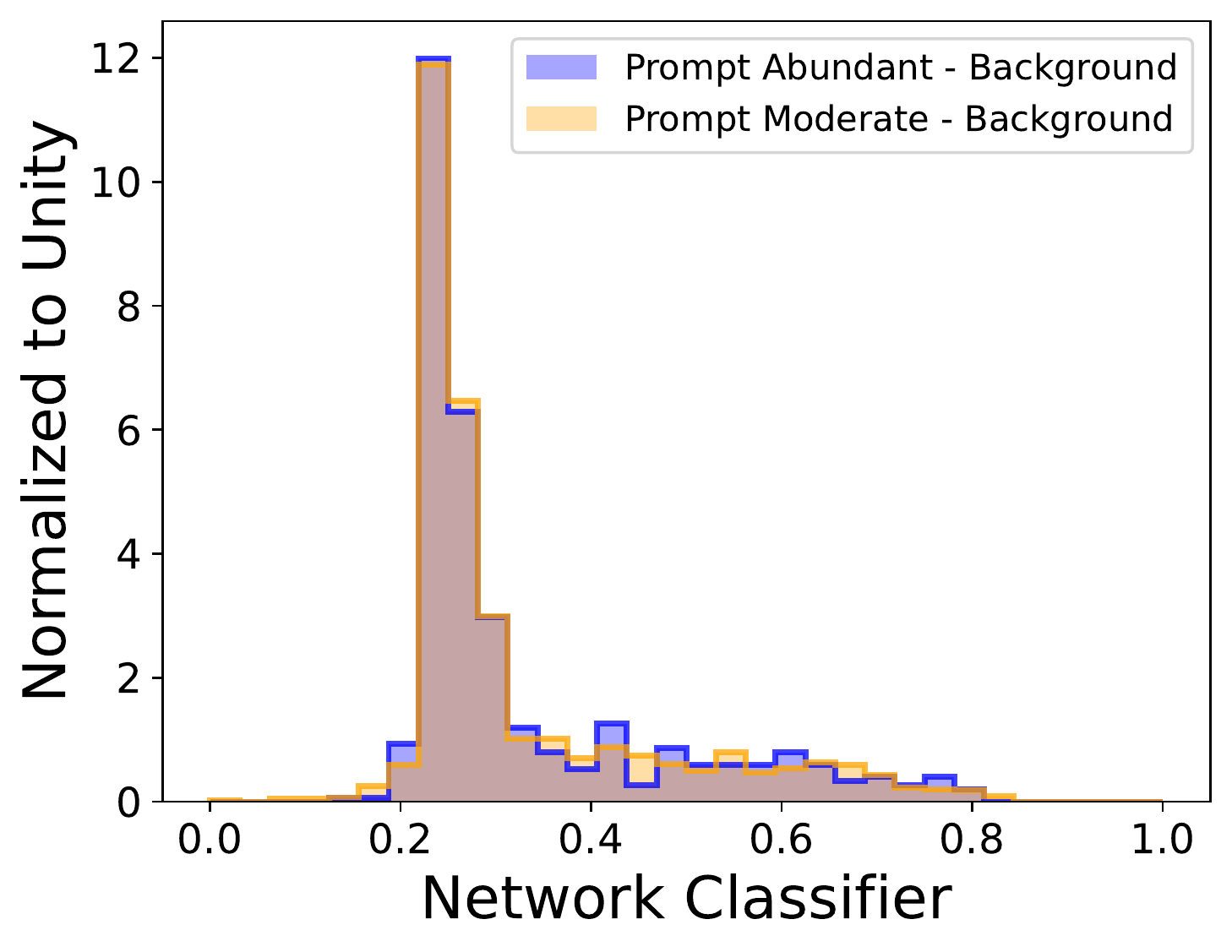}
\caption {(MG5+Pythia+Delphes) Similar to Fig~\ref{fig:sim_pT_mixed_samples}, but demonstrating that network output distributions for each class match across mixed samples.
}
\label{fig:sim_nn_mixed_samples}
\end{figure}

 Using the simulated dataset, we compute one of the features included in our models which use the CMS dataset, the summed transverse momentum of the objects in an event. We see in Fig~\ref{fig:sim_pT_mixed_samples} that the component distributions do approximately match across the samples for the simulated dataset. Similarly, the class components of a network classifier should be distributed the same way, regardless of which mixed sample the events were drawn from. We check this by training a PFN using the simulated dataset and looking at the distributions of the outputs, as shown in Fig~\ref{fig:sim_nn_mixed_samples}. Once again we see that the distributions depend on the class rather than the mixed sample to which events belong.

\end{document}